\documentclass[aps,prd,twocolumn,amsmath,superscriptaddress,amssymb,showpacs,floatfix,nofootinbib]{revtex4-1}

\usepackage{graphicx}
\usepackage{dcolumn}
\usepackage{xcolor}
\usepackage{bm}
\usepackage{natbib}
\usepackage{calligra}
\usepackage[T1]{fontenc}
\usepackage{egothic}
\usepackage[T1]{fontenc}
\newfont{\rsfsten}{rsfs10 scaled 1200}
\newfont{\rsfsseven}{rsfs10 scaled 1200}
\newfont{\rsfsfive}{rsfs10 scaled 1200}
\usepackage{epsfig}
\usepackage{units}
\usepackage[utf8]{inputenc}

\newcommand{\be}{\begin{equation}}
\newcommand{\ee}{\end{equation}}
\newcommand{\bea}{\begin{eqnarray}}
\newcommand{\eea}{\end{eqnarray}}

\def\lsim{\mathrel{\raise.3ex\hbox{$<$\kern-.75em\lower1ex\hbox{$\sim$}}}}
\def\gsim{\mathrel{\raise.3ex\hbox{$>$\kern-.75em\lower1ex\hbox{$\sim$}}}}

\begin{document}

\title{Features in the Spectrum of Cosmic-Ray Positrons from Pulsars}

\author{Ilias Cholis}
\author{Tanvi Karwal}
\author{Marc Kamionkowski}
\affiliation{Department of Physics and Astronomy, The Johns Hopkins University, Baltimore, Maryland, 21218, USA}

\date{\today}

\begin{abstract}
Pulsars have been invoked to explain the origin of recently
observed high-energy Galactic cosmic-ray positrons.  Since the
positron propagation distance decreases with energy, the number of
pulsars that can contribute to the observed positrons decreases
from $O(10^3)$ for positron energies $E\gtrsim10$ GeV to only a
few for $E \gtrsim 500$ GeV.  Thus, if pulsars explain these
positrons, the positron energy spectrum should become
increasingly bumpy at higher energies.  Here we present a
power-spectrum analysis that can be applied to seek such
spectral features in the energy spectrum for cosmic-ray
positrons and for the energy spectrum of the combined
electron/positron flux.  We account for uncertainties in the
pulsar distribution by generating hundreds of simulated spectra
from pulsar distributions consistent with current observational
constraints.  Although the current \textit{AMS-02} data do not
exhibit evidence for spectral features, we find that such
features would be detectable at the 2$\sigma$ lavel in $\simeq 10\%$
 of our simulations, with 20 years of \textit{AMS-02} data or three 
 years of \textit{DAMPE} measurements on the 
electron-plus-positron flux.
\end{abstract}


\maketitle


Cosmic-ray (CR) antimatter provides a probe of new phenomena at high
energies. Most antimatter CRs are produced via inelastic
collisions of regular high energy CR nuclei with the
interstellar medium(ISM) gas. The resulting stable particles
from these interactions are referred to as CR secondaries, and
the observed fluxes are well described by models
\cite{Moskalenko:2001ya, Kachelriess:2015wpa, GALPROPSite,
Strong:2015zva, Evoli:2008dv, DRAGONweb, Evoli:2011id,
Pato:2010ih}.  However, the CR positron flux, and energy
spectrum of the positron fraction $e^{+}/(e^{+} + e^{-})$,
is under-predicted above 10GeV by these models. Since energy losses 
from synchrotron emission and inverse Compton scattering are much 
more important for $e^{\pm}$ 
than nuclei, this discrepancy in the high-energy positron flux is expected to
be local; associated with the propagation of CRs in 
the local $\sim$kpc$^3$ volume
\cite{DiBernardo:2010is} or with characteristics of CR
$e^{\pm}$ sources in the same volume.  These sources could be
local supernova remnants (SNRs) \cite{Blasi:2009hv,
Mertsch:2009ph, Ahlers:2009ae, Blasi:2009bd, Kawanaka:2010uj, Fujita:2009wk,
Cholis:2013lwa, Mertsch:2014poa, DiMauro:2014iia, Kohri:2015mga}, local pulsars
\cite{1987ICRC....2...92H, 1995PhRvD..52.3265A, 1995A&A...294L..41A, Hooper:2008kg, Yuksel:2008rf, Profumo:2008ms,
Malyshev:2009tw, Kawanaka:2009dk, Grasso:2009ma, Linden:2013mqa, Cholis:2013psa,
Yuan:2013eja, Yin:2013vaa} or particle dark matter (DM)
\cite{Bergstrom:2008gr, Cirelli:2008jk, Cholis:2008hb,
Cirelli:2008pk, Nelson:2008hj, ArkaniHamed:2008qn,
Cholis:2008qq, Cholis:2008wq, Harnik:2008uu, Fox:2008kb,
Pospelov:2008jd, MarchRussell:2008tu, Chang:2011xn,
Cholis:2013psa,  Dienes:2013xff, Finkbeiner:2007kk,
Kopp:2013eka, Dev:2013hka}.

A number of observations suggest that SNRs are
the primary source of Galactic CR nuclei with energies up to
$O(100)$TeV.  Yet, SNRs can explain the
positron fraction only if the metallicities of environments
of recent SNRs within $\simeq$kpc are different from those averaged
within 10kpc \cite{Cholis:2013lwa, Mertsch:2014poa, Cholis:2017qlb, Tomassetti:2017izg}. 
DM explanations for the CR positron excess are constrained by
cosmic-microwave-background data \cite{Slatyer:2009yq, Evoli:2012zz,
Madhavacheril:2013cna,  Ade:2015xua, Slatyer:2015jla,
Poulin:2016nat} and $\gamma$-rays \cite{Tavakoli:2013zva,
Geringer-Sameth:2014qqa, Ackermann:2015zua}, but parts of the
parameter space are still available.  Pulsars
are a natural source of hard CR $e^{\pm}$ injection into the
ISM.  However, at the highest observed energies, $\gsim 500$GeV,
only a few very local sources, including Geminga, Monogem, and
Vela, would dominate the CR flux.  With recent observations from
HAWC \cite{Abeysekara:2017hyn, Abeysekara:2017old} and Milagro
\cite{Abdo:2009ku} of $\gsim 10$TeV $\gamma$-ray halos
at $O(10)$pc around Geminga and Monogem, we now have strong
indications that CR $e^{\pm}$ exit the surrounding pulsar wind
nebulae (PWNe) \cite{Hooper:2017gtd}, with additional
implications for both pulsar searches
\cite{Linden:2017vvb} and the TeV emission observed by HESS
\cite{Abramowski:2016mir} towards the Galactic center
\cite{Hooper:2017rzt}.
 
\begin{figure}
\begin{centering}
\hspace{-0.6cm}
\includegraphics[width=3.60in,angle=0]{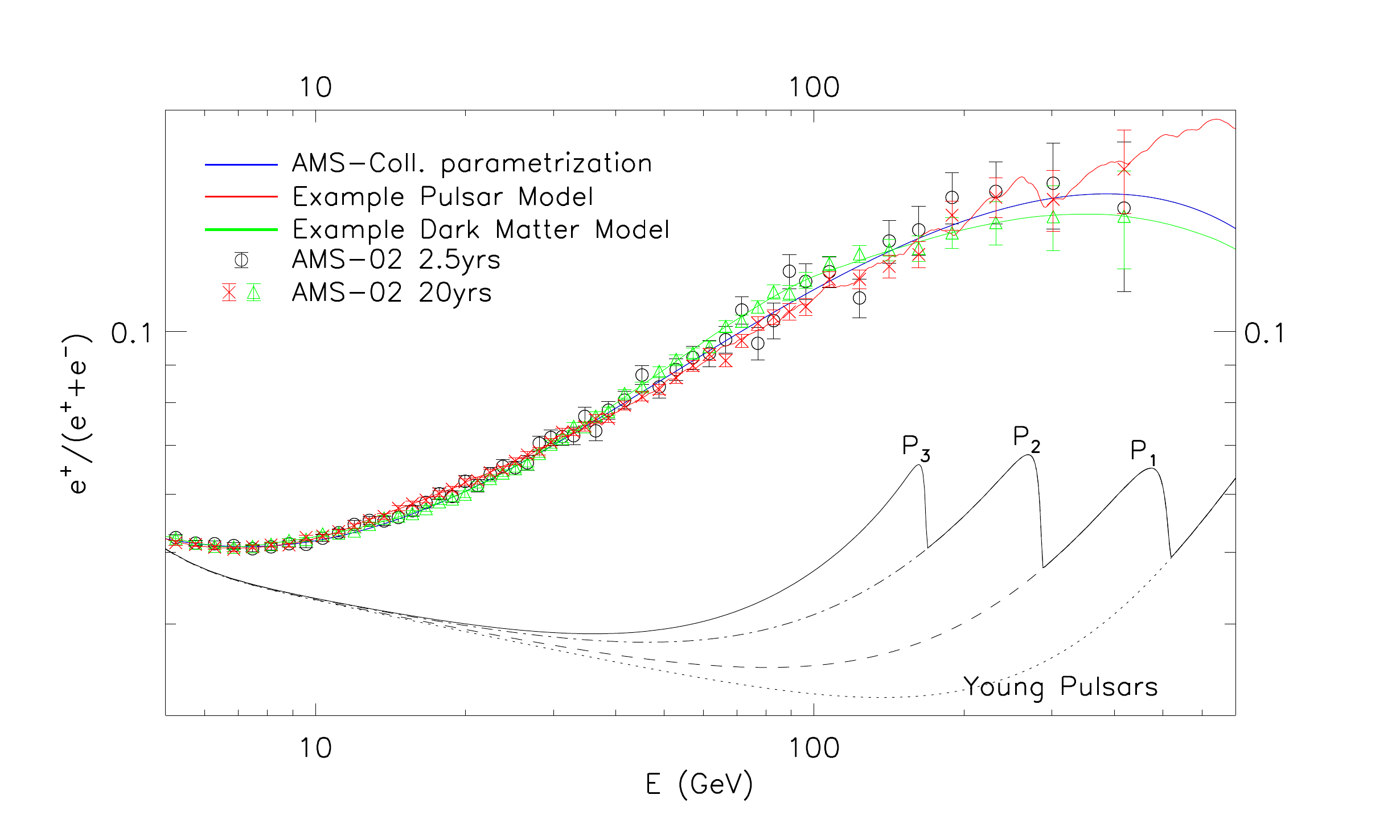}
\end{centering}
\vspace{-0.7cm}
\caption{The \textit{AMS-02} positron fraction measurement
\cite{Accardo:2014lma} and two examples of models that
fit it well. The red line is from the contribution of many Milky
Way pulsars, while the green line is from a sample DM model. 
The DM spectrum is genuinely smooth,
while the pulsar spectrum shows evidence of contributions from 
individual sources at high energies. The curves labeled 
P$_{1}$, P$_{2}$ and P$_{3}$, illustrate schematically the contribution 
from individual pulsars, at distances of 0.66, 0.97 and 1.7 kpc and ages 
of 240, 430 and 740 kyrs, respectively. 
The dotted line shows the 
contribution from pulsars with ages $\leq$150kyrs. We include 
a measurement of each model after 20yr with \textit{AMS-02}. 
We would not be able to separate them through a fit to the 
spectrum.  We include the AMS-Collaboration
parametrization \cite{Accardo:2014lma}.}
\vspace{-0.3cm}
\label{fig:PSRS_vs_DM}
\end{figure}

Pulsars are born in the Milky Way at a rate of $\simeq$1 per
century \cite{1999MNRAS.302..693D, Vranesevic:2003tp,
FaucherGiguere:2005ny, Lorimer:2006qs, Keane:2008jj}.  Thus
only one new pulsar every $10^{3}$ years is born within 
4-kpc distance that $\gtrsim10$GeV positrons can travel.
Moreover, since the energy-loss timescale is $\sim10$Myr
for $E\gtrsim$GeV positrons, no more than $\sim10^{4}$ pulsars
can contribute positrons with energies above a few GeV.
Above 100GeV the equivalent distance drops to 2 kpc and the maximum age to
2Myr, and above 500GeV to 1kpc and 400kyr.  
Thus, as we go to higher energies, the number of candidate
pulsar sources decreases.  Given the rough maximum $e^\pm$ energy
$E_{\rm max} \sim 100\, {\rm GeV} (R/2\,{\rm kpc})^{-2}$ from a
pulsar at a distance $R$, the discreteness of the source
population shows up as spectral features in the CR spectra
\cite{Malyshev:2009tw,Grasso:2009ma}. This is illustrated schematically 
in Fig.~\ref{fig:PSRS_vs_DM}. These, moreover, cannot
be mimicked by DM (even if there are multiple DM
particles) \cite{Cholis:2009va, Dienes:2013xff}.

The red-curve in Fig.~\ref{fig:PSRS_vs_DM} illustriates the type of
spectral features nduced by
discreteness of the source population.  Shown is the
positron-fraction for a simulation of pulsars
born within 4kpc  from the Sun at a rate of 1kyr$^{-1}$. 
The amplitude of the wiggles
increases as the number of contributing sources
decreases.  We show for comparison the
prediction from an example DM model (green-line)
from \cite{Cholis:2013psa} typical of~\cite{Finkbeiner:2007kk, Cholis:2008vb,
ArkaniHamed:2008qn,Cholis:2008qq}. Both the DM and
pulsar models give good fits to the \textit{AMS-02} measurement.
Even with 20 years of data, given the combined statistical and systematic
errors~\cite{Accardo:2014lma}, \textit{AMS-02} will not distinguish the DM
model from the smoothed version of the red curve.  The red curve
may, however, be distinguished through the presence of the wiggles.

In this \emph{Letter}, we suggest a power-spectrum technique to
search for wiggles in the positron energy spectrum
induced by discreteness of the source population. 
We perform 900 simulations of the Milky Way pulsar
population accounting for the astrophysical uncertainties in
this population.  We then evaluate the prospects to detect, with
this power-spectrum analysis, pulsar-induced wiggles.  While
current data are unlikely to have sufficient sensitivity, we
find that the prospects to detect wiggles with forthcoming
data are good enough to warrant a careful analysis.

\textit{Data:} We use published \textit{AMS-02} data
\cite{Accardo:2014lma} that stem from 2.5 years 
of measurements from 5 GeV and up to 500 GeV.  We also simulate
for 20 yr assuming the same energy bins and
percentage systematic errors. We also project three years of
spectral measurements of the combined $e^\pm$ flux, up to 1 TeV,
by \textit{DAMPE}. In this letter we work with binned data, but note that 
there may be benefits, in a realistic analysis, to working with the raw 
data; especially if the bin widths exceed the instrumental resolution.

\textit{Pulsar-population uncertainties:} The pulsars
contribution to the local CR spectra, has 
several uncertainties. There are uncertainties on the
neutron-star distribution in the Milky
Way~\cite{FaucherGiguere:2005ny, Lorimer:2003qc, Lorimer:2006qs}
and their birth rate \cite{1999MNRAS.302..693D,
Vranesevic:2003tp, FaucherGiguere:2005ny, Lorimer:2006qs,
Keane:2008jj}.  For the spatial distribution we follow
Ref.~\cite{Lorimer:2006qs}, which relied on data from
Ref.~\cite{Manchester:2001fp}, and take a birth rate of 
1/century. Details may be found in Appendix~\ref{appA}.  There
are also uncertainties regarding the neutron stars' initial
spin-down power $\dot{E}_{0}$, braking index $\kappa$, and
spin-down timescale $\tau_{0}$, that all relate to the time 
$t$-dependence of the spin-down power,
\begin{equation}
\dot{E}(t) = \dot{E_{0}}  \bigg(1 + \frac{t}{\tau_{0}}  \bigg)^{-\frac{\kappa+1}{\kappa-1}}.
\label{eq:SpinDown}
\end{equation}
We assume that $\dot{E}_{0}$ follows a distribution
$f(\dot{E}_{0})$, which we vary, in addition to varying $\kappa$
and $\tau_{0}$, ensuring that no pulsar has a spin-down power
higher than the recorded ones \cite{Manchester:2004bp,
ATNFSite}, with details found in Appendix~\ref{appB}.

Only a small fraction $\eta$, of the spin-down power can go to
 injected CR $e^{\pm}$ into the ISM. The CR $e^{\pm}$ before entering 
the ISM may be accelerated by the surrounding PWN and for
younger pulsars the SNR shock front further out; this leads to
significant uncertainties in the CR injection spectra. For each
pulsar we assume a unique $\eta$ and energy spectrum
$\frac{dN}{dE}_{e^{\pm}} \propto E^{-n} \, Exp
\{-E/E_{\textrm{cut}}\}$, where $\eta$ and $n$ are described by
equivalent distributions (i.e. no two pulsars in our simulations
have identical  $\eta$ or $n$).  We test different variations on
these distributions (details found in Appendix~\ref{appC}). 
Finally as CR $e^{\pm}$ enter
the ISM, they must propagate to the Earth where they are
observed. There are uncertainties regarding the CR diffusion,
energy-losses and the impact of the time-evolving Heliosphere.
We have different assumptions to model the propagation through
the ISM, using \cite{Malyshev:2009tw, Cholis:2015gna}, while we account for the
uncertainties of the propagation inside the Heliosphere
(i.e. Solar Modulation~\cite{1968ApJ...154.1011G}) by
marginalizing over them following \cite{Cholis:2015gna} and
\cite{Cholis:2017qlb}.  For the time-evolution of the
heliospheric magnetic field we use information from
Ref.~\cite{ACESite, WSOSite} (relevant details in
Appendix~\ref{appD}). 

Given a spatial distribution and pulsar birth rate, a
distribution $f(\dot{E}_{0})$, choices for  $\kappa$ and $\tau_{0}$,
distributions on the fraction of spin-down power that goes into 
ISM CR $e^{\pm}$ $g(\eta)$, distribution $h(n)$ on the
injection index $n$, and choice of ISM propagation models, we
generate a population of Milky Way pulsars that are within 4
kpc. To understand the impact of these uncertainties on the
prospects to detect fluctuations in the positron energy
spectrum, we produce 900 astrophysical realizations. Each one
has a unique combination of the above ingredients while still
consistent with pulsar population studies
\cite{FaucherGiguere:2005ny} and data on CR propagation in the
ISM and the Heliosphere \cite{Trotta:2010mx}.

\textit{Technique:} We fit the pulsar contribution to the
\textit{AMS-02} positron-fraction (containing 51 data points 
between 5 and 500 GeV)\footnote{In
principle, the analysis may be done with the
positron flux instead, since in the positron fraction,
the fluctuation amplitude is suppressed given the fluctuations
in the electron spectrum.  Still, (a) most
electrons in the relevant energy range are not from pulsars,
so the suppression is small; (b) most publicly available
current results are provided in terms of the positron fraction;
and (c) some systematic effects that might introduce artificial
fluctuations may be canceled out by working with the positron
fraction.} by allowing for an
additional normalization on the  $e^{\pm}$ pulsar flux
and by marginalizing over the uncertainties of primary and
secondary CRs and Solar Modulation (leading to five fitting parameters).  
That alone constrains a significant fraction of the pulsar astrophysical 
realizations, if they are to explain the positron fraction.  We leave that
discussion for subsequent work \cite{Cholis:2018izy}. Of the 900 pulsar  
astrophysical realizations, only 172 fit the positron fraction
within $3 \sigma$ from a prediction of 1 per degree of freedom 
i.e. with a total $\chi^{2} \leq$ 64.2 for 51-5=46 d.o.f. .
For the remainder of this analysis, we use those pulsar
astrophysical realizations, one of which is shown in
Fig.~\ref{fig:PSRS_vs_DM}. Our results are not sensitive to
the exact threshold that we place on the $\chi^{2}$ of the fit. 

For each of the remaining 172 pulsar astrophysical realizations,
we generate 10 observational realizations (i.e. add noise
following the binning and errors of
Ref.~\cite{Accardo:2014lma}); this can generate artificial
fluctuations that mask the wiggles we seek.  We then subtract from
each observational realization the smoothed spectrum
and evaluate the power-spectral density (PSD) of the residual spectrum.
Since we do not know the true underlying astrophysical spectrum,
we calculate for each realization the smoothed spectrum by
convolving with a gaussian whose width increases with
energy. This removes power in large scales in energy
(low modes in the power spectrum) including contributions from
instrument systematics as misestimates of the instrument
efficiency or CR contamination. Yet, systematic artifacts in
a small number of energy bins could still induce 
smaller-scale fluctuations that we seek.
 
To evaluate the PSD on the residual positron spectrum,
we take the ``time'' parameter to be $ln(E/\textrm{GeV})$ which
we assume to be measured in equal intervals. This is to a very
good approximation true in the energy range 5--150 GeV, with
higher energies having energy bins at larger separations. In
our calculations we assume a logarithmic energy binning of
$ln(E_{i}/\textrm{GeV}) \equiv x_{i} = x_{0} + a \cdot i$, with
$x_{0} =1.6571$ (5.24 GeV) and $a=0.063$.  When comparing to
current data, we go up to $i=59$ (215 GeV) while in our
20-year forecast we go up to $i=65$ (315 GeV).  We calculate the
PSDs for each of the 172$\times$10 observational realizations.
Given the noise, there is scatter on the PSDs of the 10
observational realizations coming from the same underlying
astrophysical realization.

To model the effect of observational scatter on the PSD, we use the
\textit{AMS-02} smooth parametrization that fits well the data
after 2.5 years and then produce 200 observational realizations
of it.  We then calculate the 200 PSDs on the residual.  Those
200 observational realizations of the \textit{AMS-02} smooth
parametrization provide the scatter on the PSD due to noise. 
For every one of the 60 (66) modes for the current (20 yr) data,
we rank the 200 coefficients. We do not expect any
correlations between modes. We use the 68$\%$ ranges to derive
the $1 \sigma$ error-bars per mode.  We then determine a
$\chi^{2}$ fit on each of the PSDs. Among the 200 observational
realizations of the \textit{AMS-02} smooth parametrization there
is a median in terms of its $\chi^{2}$ and $68\%$, $95\%$ and
$99\%$ ranges. We use these ranges to \textit{compare} the PSDs
from physical models (pulsars or DM) and the PSD from
noise. These ranges provide a measure of the scatter due to
noise and whether pulsars can give a PSD signal above the noise.
 
\begin{figure*}
\begin{centering}
\hspace{0.35cm}
\includegraphics[width=3.4in,angle=0]{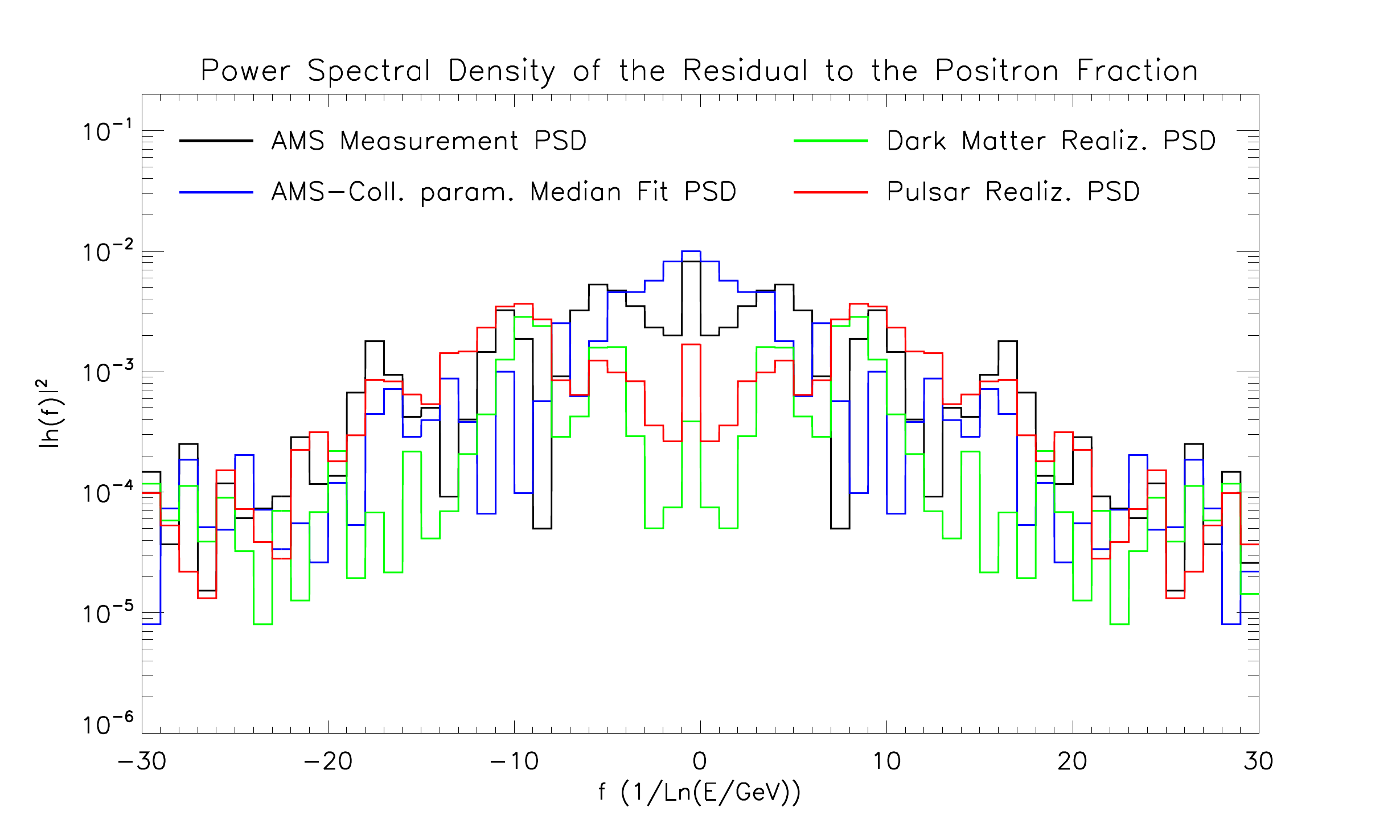}
\hspace{-0.05cm}
\includegraphics[width=3.4in,angle=0]{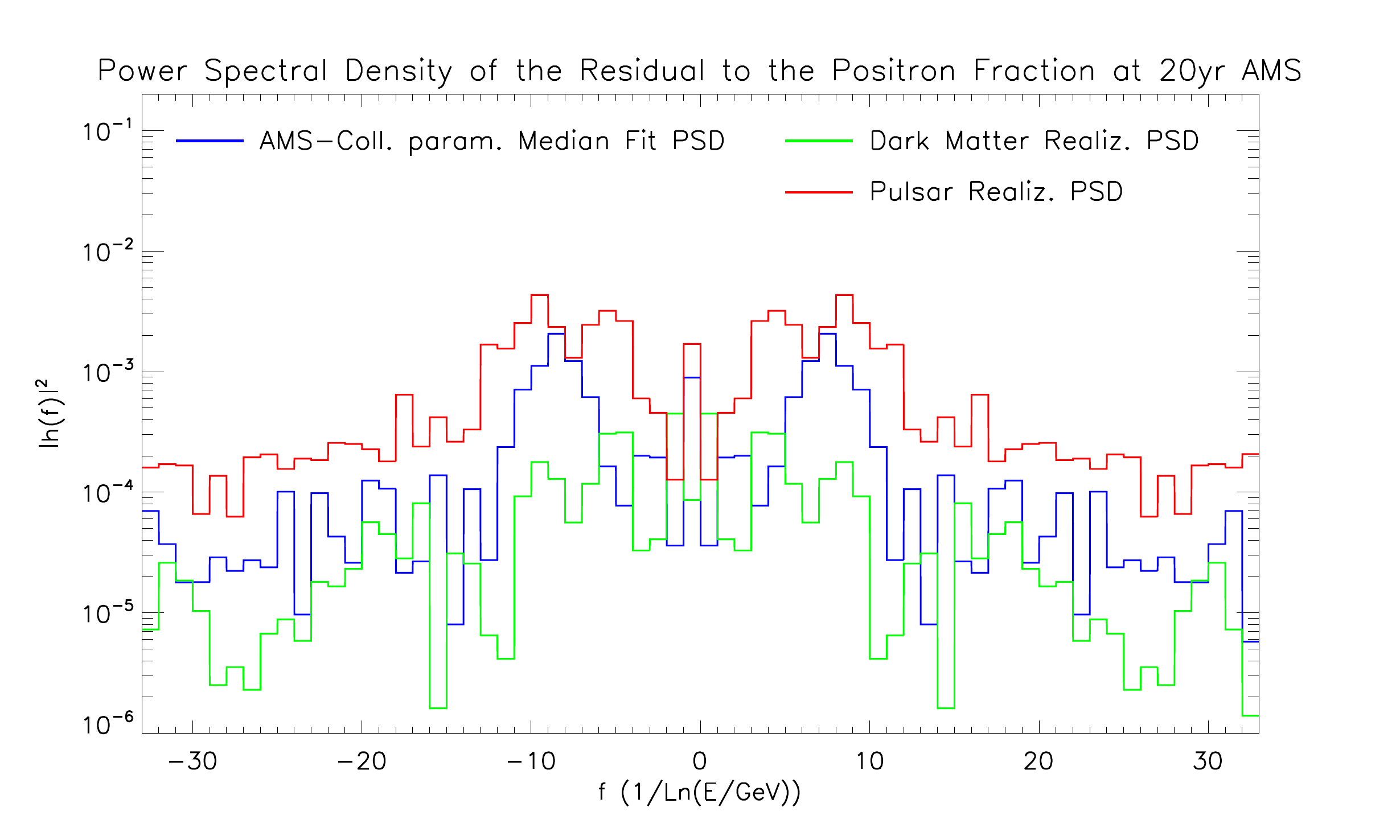}
\end{centering}
\vspace{-0.6cm}
\caption{The PSD of the residual to the positron fraction from \textit{AMS-02}. \textit{Left}: current state 
with the black line giving the measurement, the blue line the noise realization with median fit and the red and 
green lines a pulsars and a DM realization. Since pulsars have spectral features 
(shown in Fig.~\ref{fig:PSRS_vs_DM}), there is more power at the high modes of the PSD compared to the 
smooth DM realization. \textit{Right}: Same after 20 yr of
\textit{AMS-02} observations. The red line is calculated from one pulsar realization that is among the $\sim$10$\%$
of all our observation realizations, which give a signal in the PSD detectable at $\geq 2 \sigma$.} 
\label{fig:PSD_Now_and_Future}
\end{figure*}

\textit{Results:} In the left panel of Fig.~\ref{fig:PSD_Now_and_Future}
we show the PSD of the measured \textit{AMS-02} positron
fraction from the \textit{AMS-02} smooth parametrization (black
line).  The PSD of the noise realization with the median fit is
given by the blue line.  We also show the PSDs for the pulsar
astrophysical realization and the DM model
of Fig.~\ref{fig:PSRS_vs_DM} adding noise (red and green lines,
respectively).  As seen there, pulsars produce small-scale
(small in $ln(E/\textrm{GeV})$) variations that lead to
extra power in the PSD at high modes (large $f =
1/ln(E/\textrm{GeV})$). The exact DM phenomenology (i.e mass
annihilation/decay channel) can only affect the lower modes. DM
models with one very evident and sharp spectral feature that
could add some power are already 
excluded in Ref.~\cite{Bergstrom:2013jra, Ibarra:2013zia}. The
difference between the red and the green PSDs on the residual
positron fraction is what we are interested in. With 20 years of
data the situation improves, as shown in the right
panel of Fig.~\ref{fig:PSD_Now_and_Future}, for a
10$\%$ chance to see these fluctuations if pulsars are
responsible for high-energy CR positrons.

\begin{figure*}
\begin{centering}
\includegraphics[width=3.4in,angle=0]{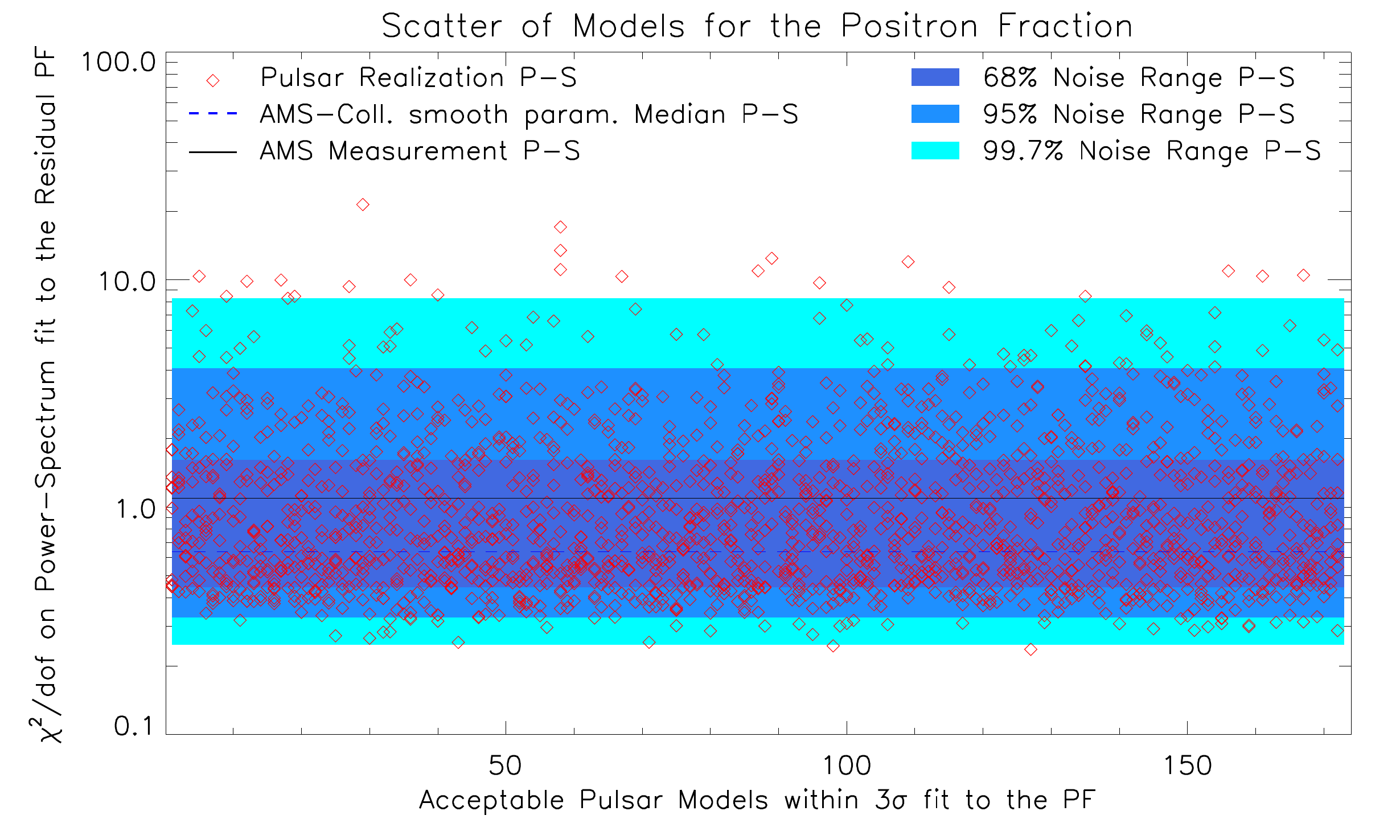}
\hspace{0.0cm}
\includegraphics[width=3.4in,angle=0]{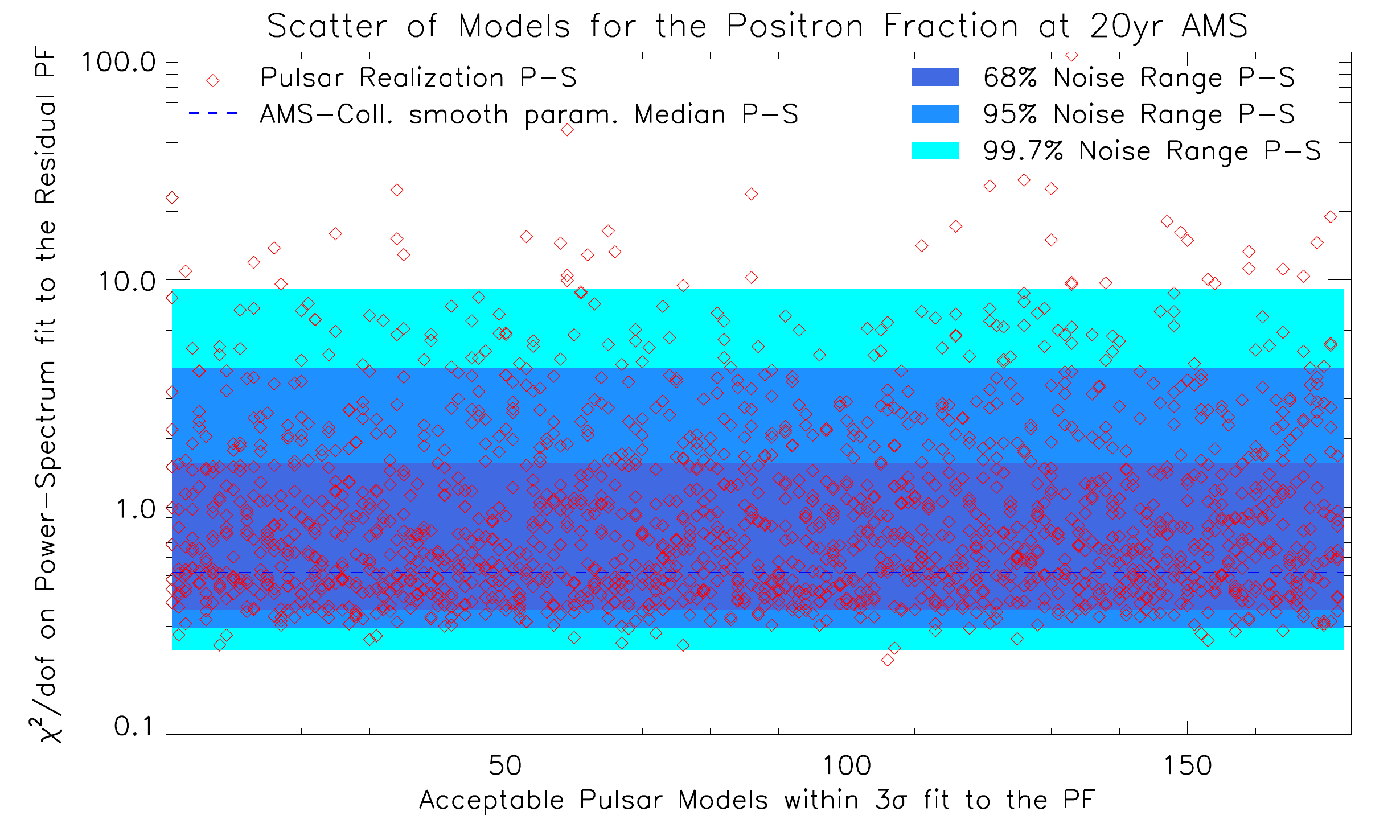}
\end{centering}
\vspace{-0.2cm}
\caption{The scatter of 10 observational realizations in the PSD $\chi^{2}$/dof for each of the 172 
pulsar astrophysical realizations (red diamonds). The blue bands 
include the noise ranges for the PSD $\chi^{2}$. \textit{Left}: Current, with the black line giving the  PSD of the measurement, showing no evidence for features. \textit{Right}: After 20 years of data, $\sim10 \%$ of the 
pulsar realizations will provide (at 2$\sigma$) detectable fluctuations.}
\vspace{-0.3cm}
\label{fig:PSFit_vs_PFFir_Now_and_Future}
\end{figure*}

In Fig.~\ref{fig:PSFit_vs_PFFir_Now_and_Future} we show for all
172 pulsar astrophysical realizations and for each of the 10
observational realizations the PSD $\chi^{2}$-distribution (red
diamonds along y-axis).  Each pulsar astrophysical realization
is in a different position on the $x$-axis; ranked starting with the
model that fits best the positron fraction spectrum. 
Our calculation of the fit of the observed
\textit{AMS-02} PSD on the residual positron fraction is given
by the black line. All diamonds, and the black line
are to be compared to the three blue bands that represent the
$68\%$, $95\%$ and $99.7\%$ ranges of the noise.  We find
that with current data 1.5$\%$ (12.5$\%$) of the 172$\times 10$
observation realizations lie outside the $99.7 \%$ ($95 \%$) band
(left panel of  Figure~\ref{fig:PSFit_vs_PFFir_Now_and_Future}).
This information is also given in Table~\ref{tab:PSTab}.  
Since we have ranked our astrophysical realization models on the 
$x$-axis by their fit to
the positron fraction, Fig.~\ref{fig:PSFit_vs_PFFir_Now_and_Future} 
also shows that there is no clear correlation between models
that provide a poor fit to the smoothed energy spectrum and
models that provide a poor fit to the power-spectrum.  Also
since the data PSD sits well within the 68$\%$ band of the
noise, there is no  indication yet that there is a deviation
from a smooth spectrum;  however, this null
result cannot yet distinguish between different scenarios.

After 20 years of observations, and using information on the
positron spectrum up to 315 GeV, the situation becomes more
promising. Then, about 2.5$\%$ (10$\%$) of the observational
realizations sit within the $99.7 \%$ ($95 \%$) noise bands, as
shown in the right panel of
Fig.~\ref{fig:PSFit_vs_PFFir_Now_and_Future} and in
Table~\ref{tab:PSTab}, with further details found in Appendix~\ref{appE}.
\begin{table}[t]
    \begin{tabular}{cccccc}
         \hline
           Experiment & E-range & $\#$frq. & $\%$exc. f. & $\%$exc. f. & $\%$exc. f. \\
                  & (GeV) & modes & (95$\%$) & (99$\%$)  & (99.7$\%$)  \\
            \hline \hline
            \textit{AMS-02} (2.5 yr) & 5.2-215 & $\pm 30$ & 12.5 & 3.5 & 1.5 \\
            \textit{AMS-02} (20 yr) & 5.2-315 &  $\pm 33$ & 10 & 6 & 2.5 \\
            \textit{DAMPE} (3 yr) &  25-640 &  $\pm 19$ & 12 & 7 & 4\\
            \hline \hline 
        \end{tabular}
\caption{The potential to observe power from small-scale features to the residual positron 
fraction (for \textit{AMS-02}) or $e^{+}+e^{-}$ flux (for \textit{DAMPE}).  We give the relevant energy range to be used, the $\#$ of modes ($\pm$ 1/2 the number of logarithmically spaced E-bins). " $\%$exc. f." gives the $\%$
of realizations that fall outside the 95$\%$, 99$\%$ and 99.7$\%$ noise ranges.} 
\vspace{-0.7cm}
\label{tab:PSTab}
\end{table}

\textit{DAMPE} \cite{TheDAMPE:2017dtc, Ambrosi:2017wek} and \textit{CALET}
\cite{2015JPhCS.632a2023A, Adriani:2017efm} are now measuring the total CR $e^{+}
+ e^{-}$ flux up to several TeV.  We forecast the prospects
to probe a PSD signal from pulsars to higher energies where
fewer pulsars contribute to the signal.  Using the
expected flux measurement between 25 and 640 GeV we find that 
38 logarithmically equally spaced energy bins provides us
with a good sensitivity to the presence of
features.\footnote{Any further optimizations should be left to
the collaborations.}  Of the 172 pulsars realizations, 53
include at least one pulsar that has similar power, age and
distance  as Geminga (PSR B0633+17) and one with similar
properties for Monogem (PSR B0656+14).  We use that subset,
since these pulsars are relevant for that range of energies but
not for the energy ranges used for the \textit{AMS-02} data. Our
findings are given in Table~\ref{tab:PSTab} suggesting that
indeed going to higher energies is necessary.

\textit{Discussion and Conclusions:} In this \emph{Letter}, we
have proposed a power-spectrum analysis to identify wiggles in
the positron energy spectrum that may arise from discreteness in
the pulsar source population, in the event that pulsars are
responsible for high-energy CR positrons.  Our basic conclusions
are that although such wiggles are likely too small to be
detectable with current data, the prospects to see such wiggles
with forthcoming data warrant the effort such
an analysis would entail.

Our estimates of the detectability of the signal rely on a
variety of uncertain ingredients in the modeling of the 
pulsar-population.  To obtain some indication of these uncertainties,
we constructed 900 simulated pulsar-population realizations
each obtained with different assumptions about the neutron-star
distribution, spin-down power characteristics and
time-evolution, the injected CR $e^{\pm}$ spectra, and propagation  
through the ISM and the heliosphere, but requiring consistency
with all observational constraints in each simulation used in
the analysis.  Thus, while our forecast of a $\sim$10$\%$ chance
to detect these wiggles is uncertain, it is, we believe, based
on realistic models.  The takeaway message is therefore that the
possibility to see something in a PSD analysis is significant
enough to warrant a search.  It is not, however, certain enough
to ascribe any strong conclusions to a null result.  With better
understanding of the astrophysics in the next decade, the
forecast may become more, or less optimistic, but almost
certainly more robust. This analysis can be repeated for SNR
sources. 

The predictions of wiggles are statistical only.  We ascribe
significance to the presence of wiggles, but we do not make
predictions about specific features at specific energies 
\footnote{With a better understanding of the local ISM propagation 
conditions, by detecting   a power-spectrum signal, we will be able 
to also constrain the number of sources within local distances.}.  We
also do not ascribe the signal to any specific pulsar
(e.g., Geminga or Monogem), although our models are required to
have pulsar-populations consistent with the existence of these pulsars.  
Also, we emphasize that we simply estimate the sensitivity of current 
measurements to a power-spectrum-based wiggle search.  We do 
hope, however, that this work motivates collaborations like 
\textit{AMS-02} and at higher energies \textit{DAMPE} and 
\textit{CALET} to perform their own PSD analysis with their data.
 
\bigskip                  
                  
We thank Mirko Boezio, Joseph Gelfand, Ely
Kovetz, Dmitry Malyshev, and Christoph Weniger for valuable
discussions.  This work was supported by NASA Grant
No.\ NNX17AK38G, NSF Grant No.\ 0244990, and the Simons
Foundation. This research project was conducted using 
computational resources  at the Maryland Advanced Research 
Computing Center (MARCC).

\bibliography{Pulsars_PF_PowerSpect}

\begin{thebibliography}{91}
\expandafter\ifx\csname natexlab\endcsname\relax\def\natexlab#1{#1}\fi
\expandafter\ifx\csname bibnamefont\endcsname\relax
  \def\bibnamefont#1{#1}\fi
\expandafter\ifx\csname bibfnamefont\endcsname\relax
  \def\bibfnamefont#1{#1}\fi
\expandafter\ifx\csname citenamefont\endcsname\relax
  \def\citenamefont#1{#1}\fi
\expandafter\ifx\csname url\endcsname\relax
  \def\url#1{\texttt{#1}}\fi
\expandafter\ifx\csname urlprefix\endcsname\relax\def\urlprefix{URL }\fi
\providecommand{\bibinfo}[2]{#2}
\providecommand{\eprint}[2][]{\url{#2}}

\bibitem[{\citenamefont{Moskalenko et~al.}(2002)\citenamefont{Moskalenko,
  Strong, Ormes, and Potgieter}}]{Moskalenko:2001ya}
\bibinfo{author}{\bibfnamefont{I.~V.} \bibnamefont{Moskalenko}},
  \bibinfo{author}{\bibfnamefont{A.~W.} \bibnamefont{Strong}},
  \bibinfo{author}{\bibfnamefont{J.~F.} \bibnamefont{Ormes}}, \bibnamefont{and}
  \bibinfo{author}{\bibfnamefont{M.~S.} \bibnamefont{Potgieter}},
  \bibinfo{journal}{Astrophys. J.} \textbf{\bibinfo{volume}{565}},
  \bibinfo{pages}{280} (\bibinfo{year}{2002}), \eprint{astro-ph/0106567}.

\bibitem[{\citenamefont{Kachelriess et~al.}(2015)\citenamefont{Kachelriess,
  Moskalenko, and Ostapchenko}}]{Kachelriess:2015wpa}
\bibinfo{author}{\bibfnamefont{M.}~\bibnamefont{Kachelriess}},
  \bibinfo{author}{\bibfnamefont{I.~V.} \bibnamefont{Moskalenko}},
  \bibnamefont{and} \bibinfo{author}{\bibfnamefont{S.~S.}
  \bibnamefont{Ostapchenko}}, \bibinfo{journal}{Astrophys. J.}
  \textbf{\bibinfo{volume}{803}}, \bibinfo{pages}{54} (\bibinfo{year}{2015}),
  \eprint{1502.04158}.

\bibitem[{\citenamefont{http://galprop.stanford.edu/.}()}]{GALPROPSite}
\bibinfo{author}{\bibnamefont{http://galprop.stanford.edu/.}}

\bibitem[{\citenamefont{Strong}(2015)}]{Strong:2015zva}
\bibinfo{author}{\bibfnamefont{A.~W.} \bibnamefont{Strong}}
  (\bibinfo{year}{2015}), \eprint{1507.05020}.

\bibitem[{\citenamefont{Evoli et~al.}(2008)\citenamefont{Evoli, Gaggero,
  Grasso, and Maccione}}]{Evoli:2008dv}
\bibinfo{author}{\bibfnamefont{C.}~\bibnamefont{Evoli}},
  \bibinfo{author}{\bibfnamefont{D.}~\bibnamefont{Gaggero}},
  \bibinfo{author}{\bibfnamefont{D.}~\bibnamefont{Grasso}}, \bibnamefont{and}
  \bibinfo{author}{\bibfnamefont{L.}~\bibnamefont{Maccione}},
  \bibinfo{journal}{JCAP} \textbf{\bibinfo{volume}{0810}}, \bibinfo{pages}{018}
  (\bibinfo{year}{2008}), \eprint{0807.4730}.

\bibitem[{\citenamefont{http://dragon.hepforge.org}()}]{DRAGONweb}
\bibinfo{author}{\bibnamefont{http://dragon.hepforge.org}}.

\bibitem[{\citenamefont{Evoli et~al.}(2012{\natexlab{a}})\citenamefont{Evoli,
  Cholis, Grasso, Maccione, and Ullio}}]{Evoli:2011id}
\bibinfo{author}{\bibfnamefont{C.}~\bibnamefont{Evoli}},
  \bibinfo{author}{\bibfnamefont{I.}~\bibnamefont{Cholis}},
  \bibinfo{author}{\bibfnamefont{D.}~\bibnamefont{Grasso}},
  \bibinfo{author}{\bibfnamefont{L.}~\bibnamefont{Maccione}}, \bibnamefont{and}
  \bibinfo{author}{\bibfnamefont{P.}~\bibnamefont{Ullio}},
  \bibinfo{journal}{Phys. Rev.} \textbf{\bibinfo{volume}{D85}},
  \bibinfo{pages}{123511} (\bibinfo{year}{2012}{\natexlab{a}}),
  \eprint{1108.0664}.

\bibitem[{\citenamefont{Pato et~al.}(2010)\citenamefont{Pato, Hooper, and
  Simet}}]{Pato:2010ih}
\bibinfo{author}{\bibfnamefont{M.}~\bibnamefont{Pato}},
  \bibinfo{author}{\bibfnamefont{D.}~\bibnamefont{Hooper}}, \bibnamefont{and}
  \bibinfo{author}{\bibfnamefont{M.}~\bibnamefont{Simet}},
  \bibinfo{journal}{JCAP} \textbf{\bibinfo{volume}{1006}}, \bibinfo{pages}{022}
  (\bibinfo{year}{2010}), \eprint{1002.3341}.

\bibitem[{\citenamefont{Di~Bernardo et~al.}(2011)\citenamefont{Di~Bernardo,
  Evoli, Gaggero, Grasso, Maccione et~al.}}]{DiBernardo:2010is}
\bibinfo{author}{\bibfnamefont{G.}~\bibnamefont{Di~Bernardo}},
  \bibinfo{author}{\bibfnamefont{C.}~\bibnamefont{Evoli}},
  \bibinfo{author}{\bibfnamefont{D.}~\bibnamefont{Gaggero}},
  \bibinfo{author}{\bibfnamefont{D.}~\bibnamefont{Grasso}},
  \bibinfo{author}{\bibfnamefont{L.}~\bibnamefont{Maccione}},
  \bibnamefont{et~al.}, \bibinfo{journal}{Astropart.Phys.}
  \textbf{\bibinfo{volume}{34}}, \bibinfo{pages}{528} (\bibinfo{year}{2011}),
  \eprint{1010.0174}.

\bibitem[{\citenamefont{Blasi}(2009)}]{Blasi:2009hv}
\bibinfo{author}{\bibfnamefont{P.}~\bibnamefont{Blasi}},
  \bibinfo{journal}{Phys. Rev. Lett.} \textbf{\bibinfo{volume}{103}},
  \bibinfo{pages}{051104} (\bibinfo{year}{2009}), \eprint{0903.2794}.

\bibitem[{\citenamefont{Mertsch and Sarkar}(2009)}]{Mertsch:2009ph}
\bibinfo{author}{\bibfnamefont{P.}~\bibnamefont{Mertsch}} \bibnamefont{and}
  \bibinfo{author}{\bibfnamefont{S.}~\bibnamefont{Sarkar}},
  \bibinfo{journal}{Phys. Rev. Lett.} \textbf{\bibinfo{volume}{103}},
  \bibinfo{pages}{081104} (\bibinfo{year}{2009}), \eprint{0905.3152}.

\bibitem[{\citenamefont{Ahlers et~al.}(2009)\citenamefont{Ahlers, Mertsch, and
  Sarkar}}]{Ahlers:2009ae}
\bibinfo{author}{\bibfnamefont{M.}~\bibnamefont{Ahlers}},
  \bibinfo{author}{\bibfnamefont{P.}~\bibnamefont{Mertsch}}, \bibnamefont{and}
  \bibinfo{author}{\bibfnamefont{S.}~\bibnamefont{Sarkar}},
  \bibinfo{journal}{Phys. Rev.} \textbf{\bibinfo{volume}{D80}},
  \bibinfo{pages}{123017} (\bibinfo{year}{2009}), \eprint{0909.4060}.

\bibitem[{\citenamefont{Blasi and Serpico}(2009)}]{Blasi:2009bd}
\bibinfo{author}{\bibfnamefont{P.}~\bibnamefont{Blasi}} \bibnamefont{and}
  \bibinfo{author}{\bibfnamefont{P.~D.} \bibnamefont{Serpico}},
  \bibinfo{journal}{Phys.Rev.Lett.} \textbf{\bibinfo{volume}{103}},
  \bibinfo{pages}{081103} (\bibinfo{year}{2009}), \eprint{0904.0871}.

\bibitem[{\citenamefont{Kawanaka et~al.}(2011)\citenamefont{Kawanaka, Ioka,
  Ohira, and Kashiyama}}]{Kawanaka:2010uj}
\bibinfo{author}{\bibfnamefont{N.}~\bibnamefont{Kawanaka}},
  \bibinfo{author}{\bibfnamefont{K.}~\bibnamefont{Ioka}},
  \bibinfo{author}{\bibfnamefont{Y.}~\bibnamefont{Ohira}}, \bibnamefont{and}
  \bibinfo{author}{\bibfnamefont{K.}~\bibnamefont{Kashiyama}},
  \bibinfo{journal}{Astrophys. J.} \textbf{\bibinfo{volume}{729}},
  \bibinfo{pages}{93} (\bibinfo{year}{2011}), \eprint{1009.1142}.

\bibitem[{\citenamefont{Fujita et~al.}(2009)\citenamefont{Fujita, Kohri,
  Yamazaki, and Ioka}}]{Fujita:2009wk}
\bibinfo{author}{\bibfnamefont{Y.}~\bibnamefont{Fujita}},
  \bibinfo{author}{\bibfnamefont{K.}~\bibnamefont{Kohri}},
  \bibinfo{author}{\bibfnamefont{R.}~\bibnamefont{Yamazaki}}, \bibnamefont{and}
  \bibinfo{author}{\bibfnamefont{K.}~\bibnamefont{Ioka}},
  \bibinfo{journal}{Phys. Rev.} \textbf{\bibinfo{volume}{D80}},
  \bibinfo{pages}{063003} (\bibinfo{year}{2009}), \eprint{0903.5298}.

\bibitem[{\citenamefont{Cholis and Hooper}(2014)}]{Cholis:2013lwa}
\bibinfo{author}{\bibfnamefont{I.}~\bibnamefont{Cholis}} \bibnamefont{and}
  \bibinfo{author}{\bibfnamefont{D.}~\bibnamefont{Hooper}},
  \bibinfo{journal}{Phys. Rev.} \textbf{\bibinfo{volume}{D89}},
  \bibinfo{pages}{043013} (\bibinfo{year}{2014}), \eprint{1312.2952}.

\bibitem[{\citenamefont{Mertsch and Sarkar}(2014)}]{Mertsch:2014poa}
\bibinfo{author}{\bibfnamefont{P.}~\bibnamefont{Mertsch}} \bibnamefont{and}
  \bibinfo{author}{\bibfnamefont{S.}~\bibnamefont{Sarkar}},
  \bibinfo{journal}{Phys. Rev.} \textbf{\bibinfo{volume}{D90}},
  \bibinfo{pages}{061301} (\bibinfo{year}{2014}), \eprint{1402.0855}.

\bibitem[{\citenamefont{Di~Mauro et~al.}(2014)\citenamefont{Di~Mauro, Donato,
  Fornengo, Lineros, and Vittino}}]{DiMauro:2014iia}
\bibinfo{author}{\bibfnamefont{M.}~\bibnamefont{Di~Mauro}},
  \bibinfo{author}{\bibfnamefont{F.}~\bibnamefont{Donato}},
  \bibinfo{author}{\bibfnamefont{N.}~\bibnamefont{Fornengo}},
  \bibinfo{author}{\bibfnamefont{R.}~\bibnamefont{Lineros}}, \bibnamefont{and}
  \bibinfo{author}{\bibfnamefont{A.}~\bibnamefont{Vittino}},
  \bibinfo{journal}{JCAP} \textbf{\bibinfo{volume}{1404}}, \bibinfo{pages}{006}
  (\bibinfo{year}{2014}), \eprint{1402.0321}.

\bibitem[{\citenamefont{Kohri et~al.}(2016)\citenamefont{Kohri, Ioka, Fujita,
  and Yamazaki}}]{Kohri:2015mga}
\bibinfo{author}{\bibfnamefont{K.}~\bibnamefont{Kohri}},
  \bibinfo{author}{\bibfnamefont{K.}~\bibnamefont{Ioka}},
  \bibinfo{author}{\bibfnamefont{Y.}~\bibnamefont{Fujita}}, \bibnamefont{and}
  \bibinfo{author}{\bibfnamefont{R.}~\bibnamefont{Yamazaki}},
  \bibinfo{journal}{PTEP} \textbf{\bibinfo{volume}{2016}},
  \bibinfo{pages}{021E01} (\bibinfo{year}{2016}), \eprint{1505.01236}.

\bibitem[{\citenamefont{{Harding} and {Ramaty}}(1987)}]{1987ICRC....2...92H}
\bibinfo{author}{\bibfnamefont{A.~K.} \bibnamefont{{Harding}}}
  \bibnamefont{and} \bibinfo{author}{\bibfnamefont{R.}~\bibnamefont{{Ramaty}}},
  \bibinfo{journal}{International Cosmic Ray Conference}
  \textbf{\bibinfo{volume}{2}}, \bibinfo{pages}{92} (\bibinfo{year}{1987}).

\bibitem[{\citenamefont{{Atoyan} et~al.}(1995)\citenamefont{{Atoyan},
  {Aharonian}, and {V{\"o}lk}}}]{1995PhRvD..52.3265A}
\bibinfo{author}{\bibfnamefont{A.~M.} \bibnamefont{{Atoyan}}},
  \bibinfo{author}{\bibfnamefont{F.~A.} \bibnamefont{{Aharonian}}},
  \bibnamefont{and} \bibinfo{author}{\bibfnamefont{H.~J.}
  \bibnamefont{{V{\"o}lk}}}, \bibinfo{journal}{\prd}
  \textbf{\bibinfo{volume}{52}}, \bibinfo{pages}{3265} (\bibinfo{year}{1995}).

\bibitem[{\citenamefont{{Aharonian} et~al.}(1995)\citenamefont{{Aharonian},
  {Atoyan}, and {Voelk}}}]{1995A&A...294L..41A}
\bibinfo{author}{\bibfnamefont{F.~A.} \bibnamefont{{Aharonian}}},
  \bibinfo{author}{\bibfnamefont{A.~M.} \bibnamefont{{Atoyan}}},
  \bibnamefont{and} \bibinfo{author}{\bibfnamefont{H.~J.}
  \bibnamefont{{Voelk}}}, \bibinfo{journal}{\aap}
  \textbf{\bibinfo{volume}{294}}, \bibinfo{pages}{L41} (\bibinfo{year}{1995}).

\bibitem[{\citenamefont{Hooper et~al.}(2009)\citenamefont{Hooper, Blasi, and
  Serpico}}]{Hooper:2008kg}
\bibinfo{author}{\bibfnamefont{D.}~\bibnamefont{Hooper}},
  \bibinfo{author}{\bibfnamefont{P.}~\bibnamefont{Blasi}}, \bibnamefont{and}
  \bibinfo{author}{\bibfnamefont{P.~D.} \bibnamefont{Serpico}},
  \bibinfo{journal}{JCAP} \textbf{\bibinfo{volume}{0901}}, \bibinfo{pages}{025}
  (\bibinfo{year}{2009}), \eprint{0810.1527}.

\bibitem[{\citenamefont{Yuksel et~al.}(2009)\citenamefont{Yuksel, Kistler, and
  Stanev}}]{Yuksel:2008rf}
\bibinfo{author}{\bibfnamefont{H.}~\bibnamefont{Yuksel}},
  \bibinfo{author}{\bibfnamefont{M.~D.} \bibnamefont{Kistler}},
  \bibnamefont{and} \bibinfo{author}{\bibfnamefont{T.}~\bibnamefont{Stanev}},
  \bibinfo{journal}{Phys. Rev. Lett.} \textbf{\bibinfo{volume}{103}},
  \bibinfo{pages}{051101} (\bibinfo{year}{2009}), \eprint{0810.2784}.

\bibitem[{\citenamefont{Profumo}(2011)}]{Profumo:2008ms}
\bibinfo{author}{\bibfnamefont{S.}~\bibnamefont{Profumo}},
  \bibinfo{journal}{Central Eur. J. Phys.} \textbf{\bibinfo{volume}{10}},
  \bibinfo{pages}{1} (\bibinfo{year}{2011}), \eprint{0812.4457}.

\bibitem[{\citenamefont{Malyshev et~al.}(2009)\citenamefont{Malyshev, Cholis,
  and Gelfand}}]{Malyshev:2009tw}
\bibinfo{author}{\bibfnamefont{D.}~\bibnamefont{Malyshev}},
  \bibinfo{author}{\bibfnamefont{I.}~\bibnamefont{Cholis}}, \bibnamefont{and}
  \bibinfo{author}{\bibfnamefont{J.}~\bibnamefont{Gelfand}},
  \bibinfo{journal}{Phys. Rev.} \textbf{\bibinfo{volume}{D80}},
  \bibinfo{pages}{063005} (\bibinfo{year}{2009}), \eprint{0903.1310}.

\bibitem[{\citenamefont{Kawanaka et~al.}(2010)\citenamefont{Kawanaka, Ioka, and
  Nojiri}}]{Kawanaka:2009dk}
\bibinfo{author}{\bibfnamefont{N.}~\bibnamefont{Kawanaka}},
  \bibinfo{author}{\bibfnamefont{K.}~\bibnamefont{Ioka}}, \bibnamefont{and}
  \bibinfo{author}{\bibfnamefont{M.~M.} \bibnamefont{Nojiri}},
  \bibinfo{journal}{Astrophys. J.} \textbf{\bibinfo{volume}{710}},
  \bibinfo{pages}{958} (\bibinfo{year}{2010}), \eprint{0903.3782}.

\bibitem[{\citenamefont{Grasso et~al.}(2009)}]{Grasso:2009ma}
\bibinfo{author}{\bibfnamefont{D.}~\bibnamefont{Grasso}} \bibnamefont{et~al.}
  (\bibinfo{collaboration}{Fermi-LAT}), \bibinfo{journal}{Astropart. Phys.}
  \textbf{\bibinfo{volume}{32}}, \bibinfo{pages}{140} (\bibinfo{year}{2009}),
  \eprint{0905.0636}.

\bibitem[{\citenamefont{Linden and Profumo}(2013)}]{Linden:2013mqa}
\bibinfo{author}{\bibfnamefont{T.}~\bibnamefont{Linden}} \bibnamefont{and}
  \bibinfo{author}{\bibfnamefont{S.}~\bibnamefont{Profumo}},
  \bibinfo{journal}{Astrophys.J.} \textbf{\bibinfo{volume}{772}},
  \bibinfo{pages}{18} (\bibinfo{year}{2013}), \eprint{1304.1791}.

\bibitem[{\citenamefont{Cholis and Hooper}(2013)}]{Cholis:2013psa}
\bibinfo{author}{\bibfnamefont{I.}~\bibnamefont{Cholis}} \bibnamefont{and}
  \bibinfo{author}{\bibfnamefont{D.}~\bibnamefont{Hooper}},
  \bibinfo{journal}{Phys. Rev.} \textbf{\bibinfo{volume}{D88}},
  \bibinfo{pages}{023013} (\bibinfo{year}{2013}), \eprint{1304.1840}.

\bibitem[{\citenamefont{Yuan et~al.}(2015)\citenamefont{Yuan, Bi, Chen, Guo,
  Lin, and Zhang}}]{Yuan:2013eja}
\bibinfo{author}{\bibfnamefont{Q.}~\bibnamefont{Yuan}},
  \bibinfo{author}{\bibfnamefont{X.-J.} \bibnamefont{Bi}},
  \bibinfo{author}{\bibfnamefont{G.-M.} \bibnamefont{Chen}},
  \bibinfo{author}{\bibfnamefont{Y.-Q.} \bibnamefont{Guo}},
  \bibinfo{author}{\bibfnamefont{S.-J.} \bibnamefont{Lin}}, \bibnamefont{and}
  \bibinfo{author}{\bibfnamefont{X.}~\bibnamefont{Zhang}},
  \bibinfo{journal}{Astropart. Phys.} \textbf{\bibinfo{volume}{60}},
  \bibinfo{pages}{1} (\bibinfo{year}{2015}), \eprint{1304.1482}.

\bibitem[{\citenamefont{Yin et~al.}(2013)\citenamefont{Yin, Yu, Yuan, and
  Bi}}]{Yin:2013vaa}
\bibinfo{author}{\bibfnamefont{P.-F.} \bibnamefont{Yin}},
  \bibinfo{author}{\bibfnamefont{Z.-H.} \bibnamefont{Yu}},
  \bibinfo{author}{\bibfnamefont{Q.}~\bibnamefont{Yuan}}, \bibnamefont{and}
  \bibinfo{author}{\bibfnamefont{X.-J.} \bibnamefont{Bi}},
  \bibinfo{journal}{Phys. Rev.} \textbf{\bibinfo{volume}{D88}},
  \bibinfo{pages}{023001} (\bibinfo{year}{2013}), \eprint{1304.4128}.

\bibitem[{\citenamefont{Bergstrom et~al.}(2008)\citenamefont{Bergstrom,
  Bringmann, and Edsjo}}]{Bergstrom:2008gr}
\bibinfo{author}{\bibfnamefont{L.}~\bibnamefont{Bergstrom}},
  \bibinfo{author}{\bibfnamefont{T.}~\bibnamefont{Bringmann}},
  \bibnamefont{and} \bibinfo{author}{\bibfnamefont{J.}~\bibnamefont{Edsjo}},
  \bibinfo{journal}{Phys. Rev.} \textbf{\bibinfo{volume}{D78}},
  \bibinfo{pages}{103520} (\bibinfo{year}{2008}), \eprint{0808.3725}.

\bibitem[{\citenamefont{Cirelli and Strumia}(2008)}]{Cirelli:2008jk}
\bibinfo{author}{\bibfnamefont{M.}~\bibnamefont{Cirelli}} \bibnamefont{and}
  \bibinfo{author}{\bibfnamefont{A.}~\bibnamefont{Strumia}},
  \bibinfo{journal}{PoS} \textbf{\bibinfo{volume}{IDM2008}},
  \bibinfo{pages}{089} (\bibinfo{year}{2008}), \eprint{0808.3867}.

\bibitem[{\citenamefont{Cholis et~al.}(2009{\natexlab{a}})\citenamefont{Cholis,
  Goodenough, Hooper, Simet, and Weiner}}]{Cholis:2008hb}
\bibinfo{author}{\bibfnamefont{I.}~\bibnamefont{Cholis}},
  \bibinfo{author}{\bibfnamefont{L.}~\bibnamefont{Goodenough}},
  \bibinfo{author}{\bibfnamefont{D.}~\bibnamefont{Hooper}},
  \bibinfo{author}{\bibfnamefont{M.}~\bibnamefont{Simet}}, \bibnamefont{and}
  \bibinfo{author}{\bibfnamefont{N.}~\bibnamefont{Weiner}},
  \bibinfo{journal}{Phys. Rev.} \textbf{\bibinfo{volume}{D80}},
  \bibinfo{pages}{123511} (\bibinfo{year}{2009}{\natexlab{a}}),
  \eprint{0809.1683}.

\bibitem[{\citenamefont{Cirelli et~al.}(2009)\citenamefont{Cirelli, Kadastik,
  Raidal, and Strumia}}]{Cirelli:2008pk}
\bibinfo{author}{\bibfnamefont{M.}~\bibnamefont{Cirelli}},
  \bibinfo{author}{\bibfnamefont{M.}~\bibnamefont{Kadastik}},
  \bibinfo{author}{\bibfnamefont{M.}~\bibnamefont{Raidal}}, \bibnamefont{and}
  \bibinfo{author}{\bibfnamefont{A.}~\bibnamefont{Strumia}},
  \bibinfo{journal}{Nucl. Phys.} \textbf{\bibinfo{volume}{B813}},
  \bibinfo{pages}{1} (\bibinfo{year}{2009}), \bibinfo{note}{[Addendum: Nucl.
  Phys.B873,530(2013)]}, \eprint{0809.2409}.

\bibitem[{\citenamefont{Nelson and Spitzer}(2010)}]{Nelson:2008hj}
\bibinfo{author}{\bibfnamefont{A.~E.} \bibnamefont{Nelson}} \bibnamefont{and}
  \bibinfo{author}{\bibfnamefont{C.}~\bibnamefont{Spitzer}},
  \bibinfo{journal}{JHEP} \textbf{\bibinfo{volume}{10}}, \bibinfo{pages}{066}
  (\bibinfo{year}{2010}), \eprint{0810.5167}.

\bibitem[{\citenamefont{Arkani-Hamed et~al.}(2009)\citenamefont{Arkani-Hamed,
  Finkbeiner, Slatyer, and Weiner}}]{ArkaniHamed:2008qn}
\bibinfo{author}{\bibfnamefont{N.}~\bibnamefont{Arkani-Hamed}},
  \bibinfo{author}{\bibfnamefont{D.~P.} \bibnamefont{Finkbeiner}},
  \bibinfo{author}{\bibfnamefont{T.~R.} \bibnamefont{Slatyer}},
  \bibnamefont{and} \bibinfo{author}{\bibfnamefont{N.}~\bibnamefont{Weiner}},
  \bibinfo{journal}{Phys. Rev.} \textbf{\bibinfo{volume}{D79}},
  \bibinfo{pages}{015014} (\bibinfo{year}{2009}), \eprint{0810.0713}.

\bibitem[{\citenamefont{Cholis et~al.}(2009{\natexlab{b}})\citenamefont{Cholis,
  Finkbeiner, Goodenough, and Weiner}}]{Cholis:2008qq}
\bibinfo{author}{\bibfnamefont{I.}~\bibnamefont{Cholis}},
  \bibinfo{author}{\bibfnamefont{D.~P.} \bibnamefont{Finkbeiner}},
  \bibinfo{author}{\bibfnamefont{L.}~\bibnamefont{Goodenough}},
  \bibnamefont{and} \bibinfo{author}{\bibfnamefont{N.}~\bibnamefont{Weiner}},
  \bibinfo{journal}{JCAP} \textbf{\bibinfo{volume}{0912}}, \bibinfo{pages}{007}
  (\bibinfo{year}{2009}{\natexlab{b}}), \eprint{0810.5344}.

\bibitem[{\citenamefont{Cholis et~al.}(2009{\natexlab{c}})\citenamefont{Cholis,
  Dobler, Finkbeiner, Goodenough, and Weiner}}]{Cholis:2008wq}
\bibinfo{author}{\bibfnamefont{I.}~\bibnamefont{Cholis}},
  \bibinfo{author}{\bibfnamefont{G.}~\bibnamefont{Dobler}},
  \bibinfo{author}{\bibfnamefont{D.~P.} \bibnamefont{Finkbeiner}},
  \bibinfo{author}{\bibfnamefont{L.}~\bibnamefont{Goodenough}},
  \bibnamefont{and} \bibinfo{author}{\bibfnamefont{N.}~\bibnamefont{Weiner}},
  \bibinfo{journal}{Phys. Rev.} \textbf{\bibinfo{volume}{D80}},
  \bibinfo{pages}{123518} (\bibinfo{year}{2009}{\natexlab{c}}),
  \eprint{0811.3641}.

\bibitem[{\citenamefont{Harnik and Kribs}(2009)}]{Harnik:2008uu}
\bibinfo{author}{\bibfnamefont{R.}~\bibnamefont{Harnik}} \bibnamefont{and}
  \bibinfo{author}{\bibfnamefont{G.~D.} \bibnamefont{Kribs}},
  \bibinfo{journal}{Phys. Rev.} \textbf{\bibinfo{volume}{D79}},
  \bibinfo{pages}{095007} (\bibinfo{year}{2009}), \eprint{0810.5557}.

\bibitem[{\citenamefont{Fox and Poppitz}(2009)}]{Fox:2008kb}
\bibinfo{author}{\bibfnamefont{P.~J.} \bibnamefont{Fox}} \bibnamefont{and}
  \bibinfo{author}{\bibfnamefont{E.}~\bibnamefont{Poppitz}},
  \bibinfo{journal}{Phys. Rev.} \textbf{\bibinfo{volume}{D79}},
  \bibinfo{pages}{083528} (\bibinfo{year}{2009}), \eprint{0811.0399}.

\bibitem[{\citenamefont{Pospelov and Ritz}(2009)}]{Pospelov:2008jd}
\bibinfo{author}{\bibfnamefont{M.}~\bibnamefont{Pospelov}} \bibnamefont{and}
  \bibinfo{author}{\bibfnamefont{A.}~\bibnamefont{Ritz}},
  \bibinfo{journal}{Phys. Lett.} \textbf{\bibinfo{volume}{B671}},
  \bibinfo{pages}{391} (\bibinfo{year}{2009}), \eprint{0810.1502}.

\bibitem[{\citenamefont{March-Russell and West}(2009)}]{MarchRussell:2008tu}
\bibinfo{author}{\bibfnamefont{J.~D.} \bibnamefont{March-Russell}}
  \bibnamefont{and} \bibinfo{author}{\bibfnamefont{S.~M.} \bibnamefont{West}},
  \bibinfo{journal}{Phys. Lett.} \textbf{\bibinfo{volume}{B676}},
  \bibinfo{pages}{133} (\bibinfo{year}{2009}), \eprint{0812.0559}.

\bibitem[{\citenamefont{Chang and Goodenough}(2011)}]{Chang:2011xn}
\bibinfo{author}{\bibfnamefont{S.}~\bibnamefont{Chang}} \bibnamefont{and}
  \bibinfo{author}{\bibfnamefont{L.}~\bibnamefont{Goodenough}},
  \bibinfo{journal}{Phys. Rev.} \textbf{\bibinfo{volume}{D84}},
  \bibinfo{pages}{023524} (\bibinfo{year}{2011}), \eprint{1105.3976}.

\bibitem[{\citenamefont{Dienes et~al.}(2013)\citenamefont{Dienes, Kumar, and
  Thomas}}]{Dienes:2013xff}
\bibinfo{author}{\bibfnamefont{K.~R.} \bibnamefont{Dienes}},
  \bibinfo{author}{\bibfnamefont{J.}~\bibnamefont{Kumar}}, \bibnamefont{and}
  \bibinfo{author}{\bibfnamefont{B.}~\bibnamefont{Thomas}},
  \bibinfo{journal}{Phys. Rev.} \textbf{\bibinfo{volume}{D88}},
  \bibinfo{pages}{103509} (\bibinfo{year}{2013}), \eprint{1306.2959}.

\bibitem[{\citenamefont{Finkbeiner and Weiner}(2007)}]{Finkbeiner:2007kk}
\bibinfo{author}{\bibfnamefont{D.~P.} \bibnamefont{Finkbeiner}}
  \bibnamefont{and} \bibinfo{author}{\bibfnamefont{N.}~\bibnamefont{Weiner}},
  \bibinfo{journal}{Phys. Rev.} \textbf{\bibinfo{volume}{D76}},
  \bibinfo{pages}{083519} (\bibinfo{year}{2007}), \eprint{astro-ph/0702587}.

\bibitem[{\citenamefont{Kopp}(2013)}]{Kopp:2013eka}
\bibinfo{author}{\bibfnamefont{J.}~\bibnamefont{Kopp}}, \bibinfo{journal}{Phys.
  Rev.} \textbf{\bibinfo{volume}{D88}}, \bibinfo{pages}{076013}
  (\bibinfo{year}{2013}), \eprint{1304.1184}.

\bibitem[{\citenamefont{Dev et~al.}(2014)\citenamefont{Dev, Ghosh, Okada, and
  Saha}}]{Dev:2013hka}
\bibinfo{author}{\bibfnamefont{P.~S.~B.} \bibnamefont{Dev}},
  \bibinfo{author}{\bibfnamefont{D.~K.} \bibnamefont{Ghosh}},
  \bibinfo{author}{\bibfnamefont{N.}~\bibnamefont{Okada}}, \bibnamefont{and}
  \bibinfo{author}{\bibfnamefont{I.}~\bibnamefont{Saha}},
  \bibinfo{journal}{Phys. Rev.} \textbf{\bibinfo{volume}{D89}},
  \bibinfo{pages}{095001} (\bibinfo{year}{2014}), \eprint{1307.6204}.

\bibitem[{\citenamefont{Cholis et~al.}(2017)\citenamefont{Cholis, Hooper, and
  Linden}}]{Cholis:2017qlb}
\bibinfo{author}{\bibfnamefont{I.}~\bibnamefont{Cholis}},
  \bibinfo{author}{\bibfnamefont{D.}~\bibnamefont{Hooper}}, \bibnamefont{and}
  \bibinfo{author}{\bibfnamefont{T.}~\bibnamefont{Linden}},
  \bibinfo{journal}{Phys. Rev.} \textbf{\bibinfo{volume}{D95}},
  \bibinfo{pages}{123007} (\bibinfo{year}{2017}), \eprint{1701.04406}.

\bibitem[{\citenamefont{Tomassetti and Oliva}(2017)}]{Tomassetti:2017izg}
\bibinfo{author}{\bibfnamefont{N.}~\bibnamefont{Tomassetti}} \bibnamefont{and}
  \bibinfo{author}{\bibfnamefont{A.}~\bibnamefont{Oliva}},
  \bibinfo{journal}{Astrophys. J.} \textbf{\bibinfo{volume}{844}},
  \bibinfo{pages}{L26} (\bibinfo{year}{2017}), \eprint{1707.06915}.

\bibitem[{\citenamefont{Slatyer et~al.}(2009)\citenamefont{Slatyer,
  Padmanabhan, and Finkbeiner}}]{Slatyer:2009yq}
\bibinfo{author}{\bibfnamefont{T.~R.} \bibnamefont{Slatyer}},
  \bibinfo{author}{\bibfnamefont{N.}~\bibnamefont{Padmanabhan}},
  \bibnamefont{and} \bibinfo{author}{\bibfnamefont{D.~P.}
  \bibnamefont{Finkbeiner}}, \bibinfo{journal}{Phys. Rev.}
  \textbf{\bibinfo{volume}{D80}}, \bibinfo{pages}{043526}
  (\bibinfo{year}{2009}), \eprint{0906.1197}.

\bibitem[{\citenamefont{Evoli et~al.}(2012{\natexlab{b}})\citenamefont{Evoli,
  Valdes, Ferrara, and Yoshida}}]{Evoli:2012zz}
\bibinfo{author}{\bibfnamefont{C.}~\bibnamefont{Evoli}},
  \bibinfo{author}{\bibfnamefont{M.}~\bibnamefont{Valdes}},
  \bibinfo{author}{\bibfnamefont{A.}~\bibnamefont{Ferrara}}, \bibnamefont{and}
  \bibinfo{author}{\bibfnamefont{N.}~\bibnamefont{Yoshida}},
  \bibinfo{journal}{Mon. Not. Roy. Astron. Soc.}
  \textbf{\bibinfo{volume}{422}}, \bibinfo{pages}{420}
  (\bibinfo{year}{2012}{\natexlab{b}}).

\bibitem[{\citenamefont{Madhavacheril et~al.}(2014)\citenamefont{Madhavacheril,
  Sehgal, and Slatyer}}]{Madhavacheril:2013cna}
\bibinfo{author}{\bibfnamefont{M.~S.} \bibnamefont{Madhavacheril}},
  \bibinfo{author}{\bibfnamefont{N.}~\bibnamefont{Sehgal}}, \bibnamefont{and}
  \bibinfo{author}{\bibfnamefont{T.~R.} \bibnamefont{Slatyer}},
  \bibinfo{journal}{Phys. Rev.} \textbf{\bibinfo{volume}{D89}},
  \bibinfo{pages}{103508} (\bibinfo{year}{2014}), \eprint{1310.3815}.

\bibitem[{\citenamefont{Ade et~al.}(2016)}]{Ade:2015xua}
\bibinfo{author}{\bibfnamefont{P.~A.~R.} \bibnamefont{Ade}}
  \bibnamefont{et~al.} (\bibinfo{collaboration}{Planck}),
  \bibinfo{journal}{Astron. Astrophys.} \textbf{\bibinfo{volume}{594}},
  \bibinfo{pages}{A13} (\bibinfo{year}{2016}), \eprint{1502.01589}.

\bibitem[{\citenamefont{Slatyer}(2016)}]{Slatyer:2015jla}
\bibinfo{author}{\bibfnamefont{T.~R.} \bibnamefont{Slatyer}},
  \bibinfo{journal}{Phys. Rev.} \textbf{\bibinfo{volume}{D93}},
  \bibinfo{pages}{023527} (\bibinfo{year}{2016}), \eprint{1506.03811}.

\bibitem[{\citenamefont{Poulin et~al.}(2016)\citenamefont{Poulin, Serpico, and
  Lesgourgues}}]{Poulin:2016nat}
\bibinfo{author}{\bibfnamefont{V.}~\bibnamefont{Poulin}},
  \bibinfo{author}{\bibfnamefont{P.~D.} \bibnamefont{Serpico}},
  \bibnamefont{and}
  \bibinfo{author}{\bibfnamefont{J.}~\bibnamefont{Lesgourgues}},
  \bibinfo{journal}{JCAP} \textbf{\bibinfo{volume}{1608}}, \bibinfo{pages}{036}
  (\bibinfo{year}{2016}), \eprint{1606.02073}.

\bibitem[{\citenamefont{Tavakoli et~al.}(2014)\citenamefont{Tavakoli, Cholis,
  Evoli, and Ullio}}]{Tavakoli:2013zva}
\bibinfo{author}{\bibfnamefont{M.}~\bibnamefont{Tavakoli}},
  \bibinfo{author}{\bibfnamefont{I.}~\bibnamefont{Cholis}},
  \bibinfo{author}{\bibfnamefont{C.}~\bibnamefont{Evoli}}, \bibnamefont{and}
  \bibinfo{author}{\bibfnamefont{P.}~\bibnamefont{Ullio}},
  \bibinfo{journal}{JCAP} \textbf{\bibinfo{volume}{1401}}, \bibinfo{pages}{017}
  (\bibinfo{year}{2014}), \eprint{1308.4135}.

\bibitem[{\citenamefont{Geringer-Sameth
  et~al.}(2015)\citenamefont{Geringer-Sameth, Koushiappas, and
  Walker}}]{Geringer-Sameth:2014qqa}
\bibinfo{author}{\bibfnamefont{A.}~\bibnamefont{Geringer-Sameth}},
  \bibinfo{author}{\bibfnamefont{S.~M.} \bibnamefont{Koushiappas}},
  \bibnamefont{and} \bibinfo{author}{\bibfnamefont{M.~G.}
  \bibnamefont{Walker}}, \bibinfo{journal}{Phys. Rev.}
  \textbf{\bibinfo{volume}{D91}}, \bibinfo{pages}{083535}
  (\bibinfo{year}{2015}), \eprint{1410.2242}.

\bibitem[{\citenamefont{Ackermann et~al.}(2015)}]{Ackermann:2015zua}
\bibinfo{author}{\bibfnamefont{M.}~\bibnamefont{Ackermann}}
  \bibnamefont{et~al.} (\bibinfo{collaboration}{Fermi-LAT}),
  \bibinfo{journal}{Phys. Rev. Lett.} \textbf{\bibinfo{volume}{115}},
  \bibinfo{pages}{231301} (\bibinfo{year}{2015}), \eprint{1503.02641}.

\bibitem[{\citenamefont{Abeysekara
  et~al.}(2017{\natexlab{a}})}]{Abeysekara:2017hyn}
\bibinfo{author}{\bibfnamefont{A.~U.} \bibnamefont{Abeysekara}}
  \bibnamefont{et~al.}, \bibinfo{journal}{Astrophys. J.}
  \textbf{\bibinfo{volume}{843}}, \bibinfo{pages}{40}
  (\bibinfo{year}{2017}{\natexlab{a}}), \eprint{1702.02992}.

\bibitem[{\citenamefont{Abeysekara
  et~al.}(2017{\natexlab{b}})}]{Abeysekara:2017old}
\bibinfo{author}{\bibfnamefont{A.~U.} \bibnamefont{Abeysekara}}
  \bibnamefont{et~al.} (\bibinfo{collaboration}{HAWC}),
  \bibinfo{journal}{Science} \textbf{\bibinfo{volume}{358}},
  \bibinfo{pages}{911} (\bibinfo{year}{2017}{\natexlab{b}}),
  \eprint{1711.06223}.

\bibitem[{\citenamefont{Abdo et~al.}(2009)}]{Abdo:2009ku}
\bibinfo{author}{\bibfnamefont{A.~A.} \bibnamefont{Abdo}} \bibnamefont{et~al.},
  \bibinfo{journal}{Astrophys. J.} \textbf{\bibinfo{volume}{700}},
  \bibinfo{pages}{L127} (\bibinfo{year}{2009}), \bibinfo{note}{[Erratum:
  Astrophys. J.703,L185(2009)]}, \eprint{0904.1018}.

\bibitem[{\citenamefont{Hooper et~al.}(2017{\natexlab{a}})\citenamefont{Hooper,
  Cholis, Linden, and Fang}}]{Hooper:2017gtd}
\bibinfo{author}{\bibfnamefont{D.}~\bibnamefont{Hooper}},
  \bibinfo{author}{\bibfnamefont{I.}~\bibnamefont{Cholis}},
  \bibinfo{author}{\bibfnamefont{T.}~\bibnamefont{Linden}}, \bibnamefont{and}
  \bibinfo{author}{\bibfnamefont{K.}~\bibnamefont{Fang}},
  \bibinfo{journal}{JCAP}  (\bibinfo{year}{2017}{\natexlab{a}}),
  \bibinfo{note}{[Phys. Rev.D96,103013(2017)]}, \eprint{1702.08436}.

\bibitem[{\citenamefont{Linden et~al.}(2017)\citenamefont{Linden, Auchettl,
  Bramante, Cholis, Fang, Hooper, Karwal, and Li}}]{Linden:2017vvb}
\bibinfo{author}{\bibfnamefont{T.}~\bibnamefont{Linden}},
  \bibinfo{author}{\bibfnamefont{K.}~\bibnamefont{Auchettl}},
  \bibinfo{author}{\bibfnamefont{J.}~\bibnamefont{Bramante}},
  \bibinfo{author}{\bibfnamefont{I.}~\bibnamefont{Cholis}},
  \bibinfo{author}{\bibfnamefont{K.}~\bibnamefont{Fang}},
  \bibinfo{author}{\bibfnamefont{D.}~\bibnamefont{Hooper}},
  \bibinfo{author}{\bibfnamefont{T.}~\bibnamefont{Karwal}}, \bibnamefont{and}
  \bibinfo{author}{\bibfnamefont{S.~W.} \bibnamefont{Li}},
  \bibinfo{journal}{Submitted to: Phys. Rev. D}  (\bibinfo{year}{2017}),
  \eprint{1703.09704}.

\bibitem[{\citenamefont{Abramowski et~al.}(2016)}]{Abramowski:2016mir}
\bibinfo{author}{\bibfnamefont{A.}~\bibnamefont{Abramowski}}
  \bibnamefont{et~al.} (\bibinfo{collaboration}{H.E.S.S.}),
  \bibinfo{journal}{Nature} \textbf{\bibinfo{volume}{531}},
  \bibinfo{pages}{476} (\bibinfo{year}{2016}), \eprint{1603.07730}.

\bibitem[{\citenamefont{Hooper et~al.}(2017{\natexlab{b}})\citenamefont{Hooper,
  Cholis, and Linden}}]{Hooper:2017rzt}
\bibinfo{author}{\bibfnamefont{D.}~\bibnamefont{Hooper}},
  \bibinfo{author}{\bibfnamefont{I.}~\bibnamefont{Cholis}}, \bibnamefont{and}
  \bibinfo{author}{\bibfnamefont{T.}~\bibnamefont{Linden}}
  (\bibinfo{year}{2017}{\natexlab{b}}), \eprint{1705.09293}.

\bibitem[{\citenamefont{{Dragicevich} et~al.}(1999)\citenamefont{{Dragicevich},
  {Blair}, and {Burman}}}]{1999MNRAS.302..693D}
\bibinfo{author}{\bibfnamefont{P.~M.} \bibnamefont{{Dragicevich}}},
  \bibinfo{author}{\bibfnamefont{D.~G.} \bibnamefont{{Blair}}},
  \bibnamefont{and} \bibinfo{author}{\bibfnamefont{R.~R.}
  \bibnamefont{{Burman}}}, \bibinfo{journal}{\mnras}
  \textbf{\bibinfo{volume}{302}}, \bibinfo{pages}{693} (\bibinfo{year}{1999}).

\bibitem[{\citenamefont{Vranesevic et~al.}(2004)}]{Vranesevic:2003tp}
\bibinfo{author}{\bibfnamefont{N.}~\bibnamefont{Vranesevic}}
  \bibnamefont{et~al.}, \bibinfo{journal}{Astrophys. J.}
  \textbf{\bibinfo{volume}{617}}, \bibinfo{pages}{L139} (\bibinfo{year}{2004}),
  \eprint{astro-ph/0310201}.

\bibitem[{\citenamefont{Faucher-Giguere and
  Kaspi}(2006)}]{FaucherGiguere:2005ny}
\bibinfo{author}{\bibfnamefont{C.-A.} \bibnamefont{Faucher-Giguere}}
  \bibnamefont{and} \bibinfo{author}{\bibfnamefont{V.~M.} \bibnamefont{Kaspi}},
  \bibinfo{journal}{Astrophys. J.} \textbf{\bibinfo{volume}{643}},
  \bibinfo{pages}{332} (\bibinfo{year}{2006}), \eprint{astro-ph/0512585}.

\bibitem[{\citenamefont{Lorimer et~al.}(2006)}]{Lorimer:2006qs}
\bibinfo{author}{\bibfnamefont{D.~R.} \bibnamefont{Lorimer}}
  \bibnamefont{et~al.}, \bibinfo{journal}{Mon. Not. Roy. Astron. Soc.}
  \textbf{\bibinfo{volume}{372}}, \bibinfo{pages}{777} (\bibinfo{year}{2006}),
  \eprint{astro-ph/0607640}.

\bibitem[{\citenamefont{Keane and Kramer}(2008)}]{Keane:2008jj}
\bibinfo{author}{\bibfnamefont{E.~F.} \bibnamefont{Keane}} \bibnamefont{and}
  \bibinfo{author}{\bibfnamefont{M.}~\bibnamefont{Kramer}},
  \bibinfo{journal}{Mon. Not. Roy. Astron. Soc.}
  \textbf{\bibinfo{volume}{391}}, \bibinfo{pages}{2009} (\bibinfo{year}{2008}),
  \eprint{0810.1512}.

\bibitem[{\citenamefont{Cholis and Weiner}(2009)}]{Cholis:2009va}
\bibinfo{author}{\bibfnamefont{I.}~\bibnamefont{Cholis}} \bibnamefont{and}
  \bibinfo{author}{\bibfnamefont{N.}~\bibnamefont{Weiner}}
  (\bibinfo{year}{2009}), \eprint{0911.4954}.

\bibitem[{\citenamefont{Cholis et~al.}(2009{\natexlab{d}})\citenamefont{Cholis,
  Goodenough, and Weiner}}]{Cholis:2008vb}
\bibinfo{author}{\bibfnamefont{I.}~\bibnamefont{Cholis}},
  \bibinfo{author}{\bibfnamefont{L.}~\bibnamefont{Goodenough}},
  \bibnamefont{and} \bibinfo{author}{\bibfnamefont{N.}~\bibnamefont{Weiner}},
  \bibinfo{journal}{Phys. Rev.} \textbf{\bibinfo{volume}{D79}},
  \bibinfo{pages}{123505} (\bibinfo{year}{2009}{\natexlab{d}}),
  \eprint{0802.2922}.

\bibitem[{\citenamefont{Accardo et~al.}(2014)}]{Accardo:2014lma}
\bibinfo{author}{\bibfnamefont{L.}~\bibnamefont{Accardo}} \bibnamefont{et~al.}
  (\bibinfo{collaboration}{AMS}), \bibinfo{journal}{Phys. Rev. Lett.}
  \textbf{\bibinfo{volume}{113}}, \bibinfo{pages}{121101}
  (\bibinfo{year}{2014}).

\bibitem[{\citenamefont{Lorimer}(2003)}]{Lorimer:2003qc}
\bibinfo{author}{\bibfnamefont{D.~R.} \bibnamefont{Lorimer}}
  (\bibinfo{year}{2003}), \bibinfo{note}{[IAU Symp.218,105(2004)]},
  \eprint{astro-ph/0308501}.

\bibitem[{\citenamefont{Manchester et~al.}(2001)}]{Manchester:2001fp}
\bibinfo{author}{\bibfnamefont{R.~N.} \bibnamefont{Manchester}}
  \bibnamefont{et~al.}, \bibinfo{journal}{Mon. Not. Roy. Astron. Soc.}
  \textbf{\bibinfo{volume}{328}}, \bibinfo{pages}{17} (\bibinfo{year}{2001}),
  \eprint{astro-ph/0106522}.

\bibitem[{\citenamefont{Manchester et~al.}(2005)\citenamefont{Manchester,
  Hobbs, Teoh, and Hobbs}}]{Manchester:2004bp}
\bibinfo{author}{\bibfnamefont{R.~N.} \bibnamefont{Manchester}},
  \bibinfo{author}{\bibfnamefont{G.~B.} \bibnamefont{Hobbs}},
  \bibinfo{author}{\bibfnamefont{A.}~\bibnamefont{Teoh}}, \bibnamefont{and}
  \bibinfo{author}{\bibfnamefont{M.}~\bibnamefont{Hobbs}},
  \bibinfo{journal}{Astron. J.} \textbf{\bibinfo{volume}{129}},
  \bibinfo{pages}{1993} (\bibinfo{year}{2005}), \eprint{astro-ph/0412641}.

\bibitem[{\citenamefont{http://www.atnf.csiro.au/research/pulsar/psrcat}()}]{ATNFSite}
\bibinfo{author}{\bibnamefont{http://www.atnf.csiro.au/research/pulsar/psrcat}}.

\bibitem[{\citenamefont{Cholis et~al.}(2016)\citenamefont{Cholis, Hooper, and
  Linden}}]{Cholis:2015gna}
\bibinfo{author}{\bibfnamefont{I.}~\bibnamefont{Cholis}},
  \bibinfo{author}{\bibfnamefont{D.}~\bibnamefont{Hooper}}, \bibnamefont{and}
  \bibinfo{author}{\bibfnamefont{T.}~\bibnamefont{Linden}},
  \bibinfo{journal}{Phys. Rev.} \textbf{\bibinfo{volume}{D93}},
  \bibinfo{pages}{043016} (\bibinfo{year}{2016}), \eprint{1511.01507}.

\bibitem[{\citenamefont{{Gleeson} and {Axford}}(1968)}]{1968ApJ...154.1011G}
\bibinfo{author}{\bibfnamefont{L.~J.} \bibnamefont{{Gleeson}}}
  \bibnamefont{and} \bibinfo{author}{\bibfnamefont{W.~I.}
  \bibnamefont{{Axford}}}, \bibinfo{journal}{\apj}
  \textbf{\bibinfo{volume}{154}}, \bibinfo{pages}{1011} (\bibinfo{year}{1968}).

\bibitem[{\citenamefont{http://www.srl.caltech.edu/ACE/ASC/}()}]{ACESite}
\bibinfo{author}{\bibnamefont{http://www.srl.caltech.edu/ACE/ASC/}}.

\bibitem[{\citenamefont{http://wso.stanford.edu/Tilts.html}()}]{WSOSite}
\bibinfo{author}{\bibnamefont{http://wso.stanford.edu/Tilts.html}}.

\bibitem[{\citenamefont{Trotta et~al.}(2011)\citenamefont{Trotta, Johannesson,
  Moskalenko, Porter, de~Austri, and Strong}}]{Trotta:2010mx}
\bibinfo{author}{\bibfnamefont{R.}~\bibnamefont{Trotta}},
  \bibinfo{author}{\bibfnamefont{G.}~\bibnamefont{Johannesson}},
  \bibinfo{author}{\bibfnamefont{I.~V.} \bibnamefont{Moskalenko}},
  \bibinfo{author}{\bibfnamefont{T.~A.} \bibnamefont{Porter}},
  \bibinfo{author}{\bibfnamefont{R.~R.} \bibnamefont{de~Austri}},
  \bibnamefont{and} \bibinfo{author}{\bibfnamefont{A.~W.}
  \bibnamefont{Strong}}, \bibinfo{journal}{Astrophys. J.}
  \textbf{\bibinfo{volume}{729}}, \bibinfo{pages}{106} (\bibinfo{year}{2011}),
  \eprint{1011.0037}.

\bibitem[{\citenamefont{Cholis et~al.}(2018)\citenamefont{Cholis, Karwal, and
  Kamionkowski}}]{Cholis:2018izy}
\bibinfo{author}{\bibfnamefont{I.}~\bibnamefont{Cholis}},
  \bibinfo{author}{\bibfnamefont{T.}~\bibnamefont{Karwal}}, \bibnamefont{and}
  \bibinfo{author}{\bibfnamefont{M.}~\bibnamefont{Kamionkowski}}
  (\bibinfo{year}{2018}), \eprint{1807.05230}.

\bibitem[{\citenamefont{Bergstrom et~al.}(2013)\citenamefont{Bergstrom,
  Bringmann, Cholis, Hooper, and Weniger}}]{Bergstrom:2013jra}
\bibinfo{author}{\bibfnamefont{L.}~\bibnamefont{Bergstrom}},
  \bibinfo{author}{\bibfnamefont{T.}~\bibnamefont{Bringmann}},
  \bibinfo{author}{\bibfnamefont{I.}~\bibnamefont{Cholis}},
  \bibinfo{author}{\bibfnamefont{D.}~\bibnamefont{Hooper}}, \bibnamefont{and}
  \bibinfo{author}{\bibfnamefont{C.}~\bibnamefont{Weniger}},
  \bibinfo{journal}{Phys.Rev.Lett.} \textbf{\bibinfo{volume}{111}},
  \bibinfo{pages}{171101} (\bibinfo{year}{2013}), \eprint{1306.3983}.

\bibitem[{\citenamefont{Ibarra et~al.}(2014)\citenamefont{Ibarra,
  Lamperstorfer, and Silk}}]{Ibarra:2013zia}
\bibinfo{author}{\bibfnamefont{A.}~\bibnamefont{Ibarra}},
  \bibinfo{author}{\bibfnamefont{A.~S.} \bibnamefont{Lamperstorfer}},
  \bibnamefont{and} \bibinfo{author}{\bibfnamefont{J.}~\bibnamefont{Silk}},
  \bibinfo{journal}{Phys. Rev.} \textbf{\bibinfo{volume}{D89}},
  \bibinfo{pages}{063539} (\bibinfo{year}{2014}), \eprint{1309.2570}.

\bibitem[{\citenamefont{Chang et~al.}(2017)}]{TheDAMPE:2017dtc}
\bibinfo{author}{\bibfnamefont{J.}~\bibnamefont{Chang}} \bibnamefont{et~al.}
  (\bibinfo{collaboration}{DAMPE}), \bibinfo{journal}{Astropart. Phys.}
  \textbf{\bibinfo{volume}{95}}, \bibinfo{pages}{6} (\bibinfo{year}{2017}),
  \eprint{1706.08453}.

\bibitem[{\citenamefont{Ambrosi et~al.}(2017)}]{Ambrosi:2017wek}
\bibinfo{author}{\bibfnamefont{G.}~\bibnamefont{Ambrosi}} \bibnamefont{et~al.}
  (\bibinfo{collaboration}{DAMPE}) (\bibinfo{year}{2017}), \eprint{1711.10981}.

\bibitem[{\citenamefont{{Adriani} et~al.}(2015)\citenamefont{{Adriani},
  {Akaike}, {Asano}, {Asaoka}, {Bagliesi}, {Bigongiari}, {Binns}, {Bonechi},
  {Bongi}, {Buckley} et~al.}}]{2015JPhCS.632a2023A}
\bibinfo{author}{\bibfnamefont{O.}~\bibnamefont{{Adriani}}},
  \bibinfo{author}{\bibfnamefont{Y.}~\bibnamefont{{Akaike}}},
  \bibinfo{author}{\bibfnamefont{K.}~\bibnamefont{{Asano}}},
  \bibinfo{author}{\bibfnamefont{Y.}~\bibnamefont{{Asaoka}}},
  \bibinfo{author}{\bibfnamefont{M.~G.} \bibnamefont{{Bagliesi}}},
  \bibinfo{author}{\bibfnamefont{G.}~\bibnamefont{{Bigongiari}}},
  \bibinfo{author}{\bibfnamefont{W.~R.} \bibnamefont{{Binns}}},
  \bibinfo{author}{\bibfnamefont{S.}~\bibnamefont{{Bonechi}}},
  \bibinfo{author}{\bibfnamefont{M.}~\bibnamefont{{Bongi}}},
  \bibinfo{author}{\bibfnamefont{J.~H.} \bibnamefont{{Buckley}}},
  \bibnamefont{et~al.}, in \emph{\bibinfo{booktitle}{Journal of Physics
  Conference Series}} (\bibinfo{year}{2015}), vol. \bibinfo{volume}{632} of
  \emph{\bibinfo{series}{Journal of Physics Conference Series}}, p.
  \bibinfo{pages}{012023}.

\bibitem[{\citenamefont{Adriani et~al.}(2017)}]{Adriani:2017efm}
\bibinfo{author}{\bibfnamefont{O.}~\bibnamefont{Adriani}} \bibnamefont{et~al.}
  (\bibinfo{collaboration}{CALET}), \bibinfo{journal}{Phys. Rev. Lett.}
  \textbf{\bibinfo{volume}{119}}, \bibinfo{pages}{181101}
  (\bibinfo{year}{2017}), \eprint{1712.01711}.

\end{thebibliography}
\bibliographystyle{apsrev}

\begin{appendix}

\section{The Neutron Star Distribution in Space and Time}
\label{appA}

Ref.~\cite{Lorimer:2006qs} suggests that pulsars are born in the
Milky Way at a rate of $1.4 \pm 0.2$ pulsars per century
\cite{Lorimer:2006qs}, although one finds a wider range of
estimates in other work \cite{1999MNRAS.302..693D, Vranesevic:2003tp, FaucherGiguere:2005ny, Keane:2008jj}.
For simplicity we assume a pulsar birth rate of one per century. 

The spatial distribution of pulsars in the Galaxy has been
investigated in
Ref.~\cite{FaucherGiguere:2005ny,Lorimer:2003qc,Lorimer:2006qs}
relying on data from the Parkes multi-beam pulsar survey at 1.4
GHz \cite{Manchester:2001fp}.  Our radial distribution of
pulsars is based on the best-fit parameters of
Ref.~\cite{Lorimer:2006qs}, given by an empirical expression for
the pulsar surface (column) density in the Galaxy, 
	\begin{equation}
		\rho(R) = A (R/R_{\odot})^B \mbox{exp} \left( -C\frac{R-R_{\odot}}{R_{\odot}} \right) \mbox{ kpc}^{-2},
		\label{eq-lorimer}
	\end{equation}
where $R$ is the Galactocentric radius and $R_{\odot} = 8.5$ kpc is the distance of the Sun from Galactic center (GC). We use the values $B = 1.9$ and $C=5.0$ given therein, normalizing $A$ such that we obtain our
assumed birth rate.
Our spatial simulations are consistent with
Ref.~\cite{Lorimer:2003qc} as shown in
Fig. 4. 

Using Eq.~\eqref{eq-lorimer}, we are lead to the following
probability distribution function for the radial distance of a
pulsar from GC,
\begin{equation}
	PDF(R, B, C) = 
	\frac{C^{B+2}}{A e^C R_{\odot}^2}
	\frac{ \rho(R) R }{ \Gamma (B+2) },
\end{equation}
where $\Gamma(x)$ is a Gamma function. 
We utilize a Laplace $z$-distribution with a characteristic
scale of 50 pc as done in Ref.~\cite{FaucherGiguere:2005ny} and
a flat angular distribution and simulate the pulsars within 4
kpc of the Sun. 
\begin{figure}
\begin{centering}
\includegraphics[width=3.45in,angle=0]{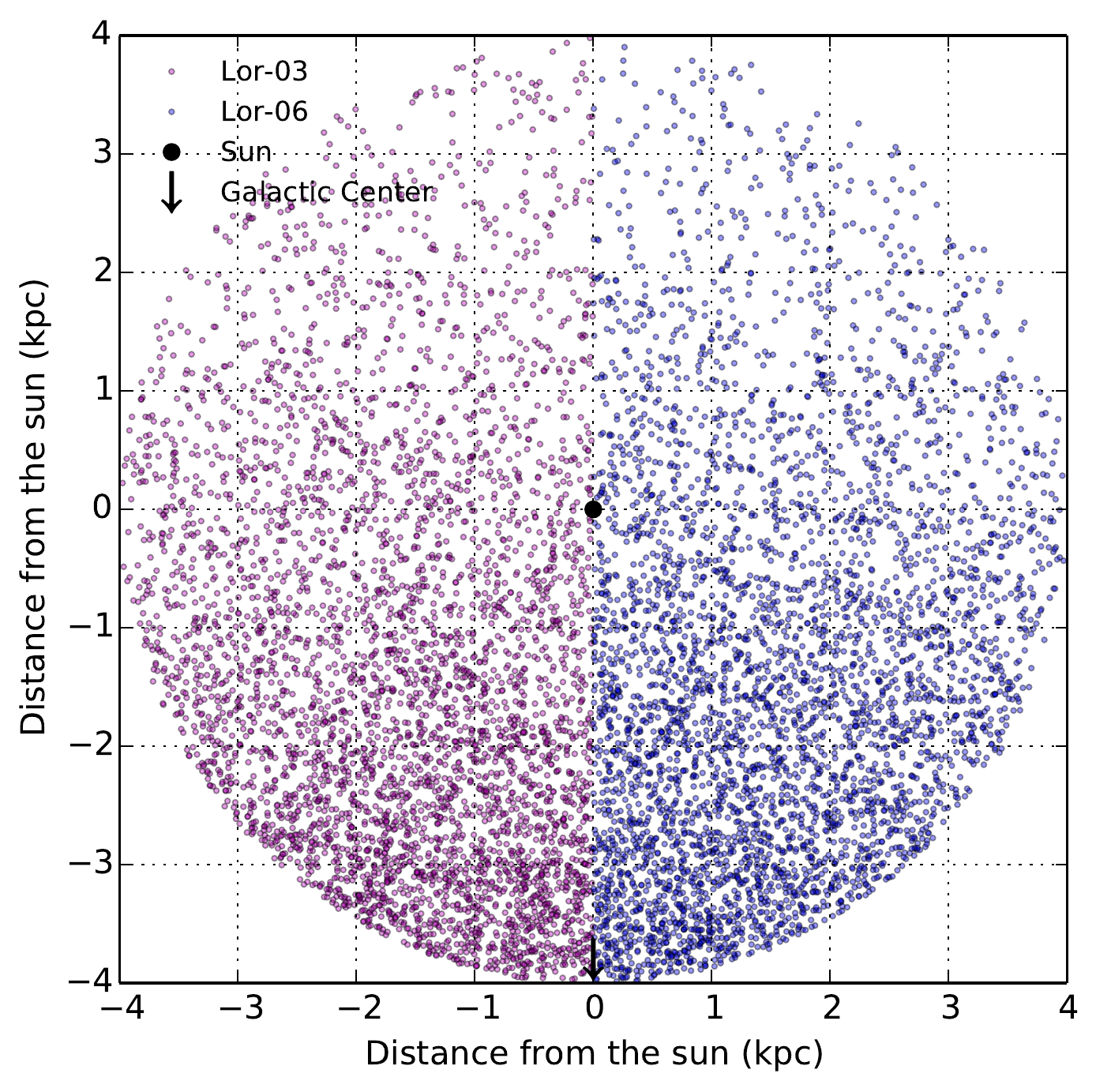}
\caption{The pulsars simulated within 4 kpc of the Sun,
projected onto the Galactic disk. The Galactic center is at $(0,
-8.5)$ kpc. The Sun is the black dot at $(0,0)$. The magenta
dots were simulated using the empirical pulsar radial
distribution curve presented in Ref.~\cite{Lorimer:2003qc}. The blue
dots were simulated using the best-fit pulsar radial
distribution curve given in Ref.~\cite{Lorimer:2006qs}. Both produce
very similar results. The number of pulsars in both simulations
were normalized such that  one per century is born in the Galaxy,
showing pulsars up to 10 Myr in age.}
\end{centering}
\label{fig:sim_psr_map_old_v_new}
\end{figure}

\section{The Neutron Stars Spin-Down Distribution Properties}
\label{appB}

Pulsar spin-down powers $\dot{E}$ are calculated using their ages and,
\begin{equation}
\dot{E}(t) = \dot{E_{0}}  \bigg(1 + \frac{t}{\tau_{0}}  \bigg)^{-\frac{\kappa+1}{\kappa-1}}.
\label{eq:SpinDown}
\end{equation}
The spin-down timescale $\tau_0$ and the braking index $\kappa$ are varied per set of simulations. 
We let $\dot{E}_0 = 10^x$ ergs/s with $x =
x_{\textrm{cutoff}} - y$ and where $y$ is taken from a
log-normal distribution. 
The log-normal distribution is generated using the parameters
$y_{\mu}$ and $y_{\sigma}$, which are the mean and
standard deviation of the underlying Gaussian distribution. We
consider four different values of $y_{\sigma} = [0.25,
0.36, 0.5, 0.75]$. Values of $x_{\textrm{cutoff}}$ and
$y_{\mu}$ are then chosen such that the distributions of
observed pulse periods and surface magnetic fields of simulated
pulsars are consistent with results presented in Fig.\ 6 of
Ref.~\cite{FaucherGiguere:2005ny}. 

Finally, to ensure that we do not produce pulsars more luminous
than the ones recorded in the ATNF catalog
\cite{Manchester:2004bp, ATNFSite}, we only consider values of
$x < x_{\textrm{max}} = 38.7$. In
Table~\ref{tab:PulsarsSim} we give all the spin-down power
distribution properties for our pulsar simulations.

\begin{table}[t]
	\centering
	\small 
	\begin{tabular}{c | *{5}{c}} 

Sim no. 	
& $\tau_{0}$ (kyr) 	
& $\kappa$ 	
& $x_{\textrm{cutoff}}$ 	
& $y_{\mu}$ 	
& $y_{\sigma}$ 	
\\
\hline
		30-59 	
		& $3.3$ 	
		& $3$ 	
		& $38.8$ 	
		& $0.25$ 	
		& $0.5$ 	
		\\
		\hline
		120-149	
		& $6$ 	
		& $3$ 	
		& $38.8$ 	
		& $0.25$ 	
		& $0.5$ 	
		\\
		150-179	
		& $3.3$ 	
		& $3$ 	
		& $38.8$ 	
		& $0.25$ 	
		& $0.5$ 	
		\\
		180-209	
		& $10$ 	
		& $3$ 	
		& $38.8$ 	
		& $0.25$ 	
		& $0.5$ 	
		\\
		210-239	
		& $3.3$ 	
		& $3$ 	
		& $39$ 	
		& $0.1$ 	
		& $0.5$ 	
		\\
		240-269	
		& $1$ 	
		& $2.5$ 	
		& $38.8$ 	
		& $0.25$ 	
		& $0.5$ 	
		\\
		270-299
		& $20$ 	
		& $3.5$ 	
		& $39$ 	
		& $0.1$ 	
		& $0.5$ 	
		\\
		300-329
		& $0.7$ 	
		& $2.5$ 	
		& $38.8$ 	
		& $0.25$ 	
		& $0.5$ 	
		\\
		330-359
		& $20$ 	
		& $3.5$ 	
		& $39.1$ 	
		& $0.0$ 	
		& $0.25$ 		
		\\
		360-389
		& $0.6$ 	
		& $2.5$ 	
		& $39.0$ 	
		& $0.1$ 	
		& $0.25$ 	
		\\
		390-419
		& $6$ 	
		& $3$ 	
		& $39.0$ 	
		& $0.1$ 	
		& $0.25$ 	
		\\
		420-449
		& $6$ 	
		& $3$ 	
		& $38.7$ 	
		& $0.5$ 	
		& $0.75$ 	
		\\
		450-479 	
		& $30$ 	
		& $3.5$ 	
		& $38.8$ 	
		& $0.25$ 	
		& $0.5$ 	
		\\
		480-509
		& $0.85$ 	
		& $2.5$ 	
		& $38.5$ 	
		& $0.6$ 	
		& $0.75$ 	 	
		\\
		510-539
		& $18$ 	
		& $3.5$ 	
		& $39.0$ 	
		& $0.0$ 	
		& $0.75$ 		
		\\
		540-569 	
		& $10$ 	
		& $3$ 	
		& $38.7$ 	
		& $0.5$ 	
		& $0.75$ 	 	
		\\
		570-599
		& $4$ 	
		& $3$ 	
		& $39.0$ 	
		& $0.0$ 	
		& $0.36$ 		
		\\
		600-629 	
		& $1$ 	
		& $2.5$ 	
		& $38.7$ 	
		& $0.5$ 	
		& $0.75$ 	
		\\
		630-659
		& $9$ 	
		& $3$ 	
		& $38.2$ 	
		& $0.4$ 	
		& $0.36$ 	
		\\
		660-689
		& $0.8$ 	
		& $2.5$ 	
		& $38.2$ 	
		& $0.4$ 	
		& $0.36$ 	
		\\
		690-719
		& $0.6$ 	
		& $2.5$ 	
		& $38.2$ 	
		& $0.4$ 	
		& $0.36$ 	
		\\
		720-749
		& $30$ 	
		& $3.5$ 	
		& $38.2$ 	
		& $0.4$ 	
		& $0.36$ 	
		\\
		750-779
		& $7$ 	
		& $3$ 	
		& $39.0$ 	
		& $0.1$ 	
		& $0.75$ 	
		\\
		780-809
		& $30$ 	
		& $3.5$ 	
		& $38.0$ 	
		& $0.5$ 	
		& $0.36$ 		
		\\
		810-839 	
		& $30$ 	
		& $3.5$ 	
		& $38.7$ 	
		& $0.5$ 	
		& $0.75$ 	 	
		\\
		840-869
		& $6$ 	
		& $3$ 	
		& $38.9$ 	
		& $0.18$ 	
		& $0.36$ 	 	
		\\
		870-899
		& $4.5$ 	
		& $3$ 	
		& $39.3$ 	
		& $0.0$ 	
		& $0.25$ 	 	
		\\
		900-929
		& $9$ 	
		& $3$ 	
		& $38.5$ 	
		& $0.5$ 	
		& $0.25$ 	 	
		\\
		930-959
		& $27$ 	
		& $3.5$ 	
		& $38.5$ 	
		& $0.3$ 	
		& $0.25$ 	
	         \\
		960-989
		& $33$ 	
		& $3.5$ 	
		& $38.0$ 	
		& $0.5$ 	
		& $0.25$ 	
		\\
		990-1019
		& $0.85$ 	
		& $2.5$ 	
		& $38.3$ 	
		& $0.5$ 	
		& $0.25$ 	
		\\

	\end{tabular}
	\caption{The assumed pulsar-simulation spin-down power
        distributions and time evolution. Simulations $\# 30-59$
        are produced based on Ref.~\cite{Lorimer:2003qc},
	while all others are on reference assumption of
        Ref.~\cite{Lorimer:2006qs}.}
	\label{tab:PulsarsSim}
\end{table}

\section{The Acceleration of CR electrons and positrons from Pulsars and Injection into the ISM}
\label{appC}

Electrons get accelerated inside the magnetosphere, produce ICS $\gamma$-rays, which in turn
in the presence of strong magnetic fields pair produce $e^{\pm}$. These $e^{\pm}$ get further accelerated inside 
the magnetosphere. In addition, electrons and positrons will then propagate outwards losing energy during
 adiabatic E-losses, but can also be accelerated in the termination shock of the pulsar(also of the SNR) and the
  ISM. 
  There is also evidence for $\gamma$-rays towards Geminga and Monogem  \cite{Abdo:2009ku, Abeysekara:2017hyn, Abeysekara:2017old},
  suggesting the presence of CR $e^{\pm}$ at 100 TeV in energy, losing a significant fraction of their energy 
  within $\simeq 10$pc. 
  Since the spin-down power drops with a time-scale of $\tau_{0} \lsim 10^{4}$ yrs, about half of the rotational
   energy will be lost before the SNR shock front stops being an efficient accelerator and well before the PWN stops
having an effect on these CRs. Given that the time for CR
$e^{\pm}$ to propagate to Earth is an order of magnitude larger
than $\tau_{0}$ we can consider their injection to the ISM
instantaneous (see Ref.~\cite{Malyshev:2009tw} for further details). 

In this work we are agnostic about the fraction $\eta$ of the spin-down power that goes into injected $e^{\pm}$.
We assume a log-normal distribution for the $\eta$ parameter, 
\begin{equation}
g(\eta) = \frac{Exp \left\{ -\frac{ \left[ - \mu +ln(-1 + \eta) \right]^{2}}{2 \sigma^2}\right\}}{\sqrt{2 \pi} (\eta -1) \sigma},
\label{eq:eta}
\end{equation}
and take three different choices for $\mu$ and $\sigma$.
These lead to three different choices for the combination of mean efficiency 
$\eta$, $\bar{\eta} = 1 + \textrm{Exp}\left\{ \mu + \frac{\sigma^2}{2} \right\}$ and logarithmic standard deviation  
$\zeta = 10^{\sigma}$: ($\bar{\eta}$, $\zeta$)  = ($4\times 10^{-3}$, 1.47) or ($10^{-3}$, 2.85) or 
($2\times 10^{-2}$, 1.29). As described in the main text, in fitting the positron fraction we allow for each 
astrophysical pulsar realization an overall normalization change in the pulsar component, that is absorbed 
into the specific values of $\bar{\eta}$. Our typical $\bar{\eta}$ is a few$\times 10^{-2}$ with a range of $2\times10^{-3} - 2\times10^{-1}$.

For the injection CR $e^{\pm}$ spectra we assume,
\begin{equation}
\frac{dN}{dE} \propto  E^{-n} Exp \left\{ -\frac{E}{E_{\textrm{cut}}}\right\},
\label{eq:InjSpect}
\end{equation}
with $n$ following a flat distribution $g(n)$ either in a narrow
range of $n \epsilon \left[1.6,1.7\right]$ or in a wider range
of $n \epsilon \left[1.4,1.9\right]$. The upper cutoff
$E_{\textrm{cut}}$ does not affect our fits to the observations,
since the highest-energy CR $e^{\pm}$ quickly lose their
energy before reaching us; we set it to $E_{\textrm{cut}} = 10$
TeV.

\section{Cosmic-Ray Propagation through the ISM and heliosphere}
\label{appD}

From the moment CRs enter into the ISM they diffuse through the
Milky Way magnetic field and suffer energy losses due to
synchrotron radiation and inverse Compton scattering. We use
five distinctive models for the ISM that 
agree with CR data including the B/C ratio, CR protons and He
\cite{Cholis:2015gna}. The 
characteristics of these five ISM models are given in Table~\ref{tab:ISMBack}. 
\begin{table}[t]
    \begin{tabular}{cccc}
         \hline
           Model & $b$ ($\times 10^{-6}$GeV$^{-1}$kyr$^{-1}$) & $D_{0}$ (pc$^2$/kyr) & $\delta$\\
            \hline \hline
            A1 &  5.05 & 123.4 & 0.33 \\
            C1 &  5.05 & 92.1 & 0.40 \\
            C2 &  8.02 & 92.1 & 0.40 \\
            C3 &  2.97 & 92.1 & 0.40 \\      
            E1 &  5.05 & 58.9 & 0.50 \\
        \hline \hline 
        \end{tabular}
\caption{The basic parameters that describe the propagations assumptions of cosmic rays in the Milky Way. Assuming isotropic and homogeneous diffusion, $D(R) = D_{0} (R/1 GV)^{\delta}$. The energy losses due to 
synchrotron radiation and inverse Compton scattering are described by $dE/dt = - b \, E^{2}$.} 
\label{tab:ISMBack}
\end{table}

\begin{figure}
\begin{centering}
\hspace{-0.6cm}
\includegraphics[width=3.60in,angle=0]{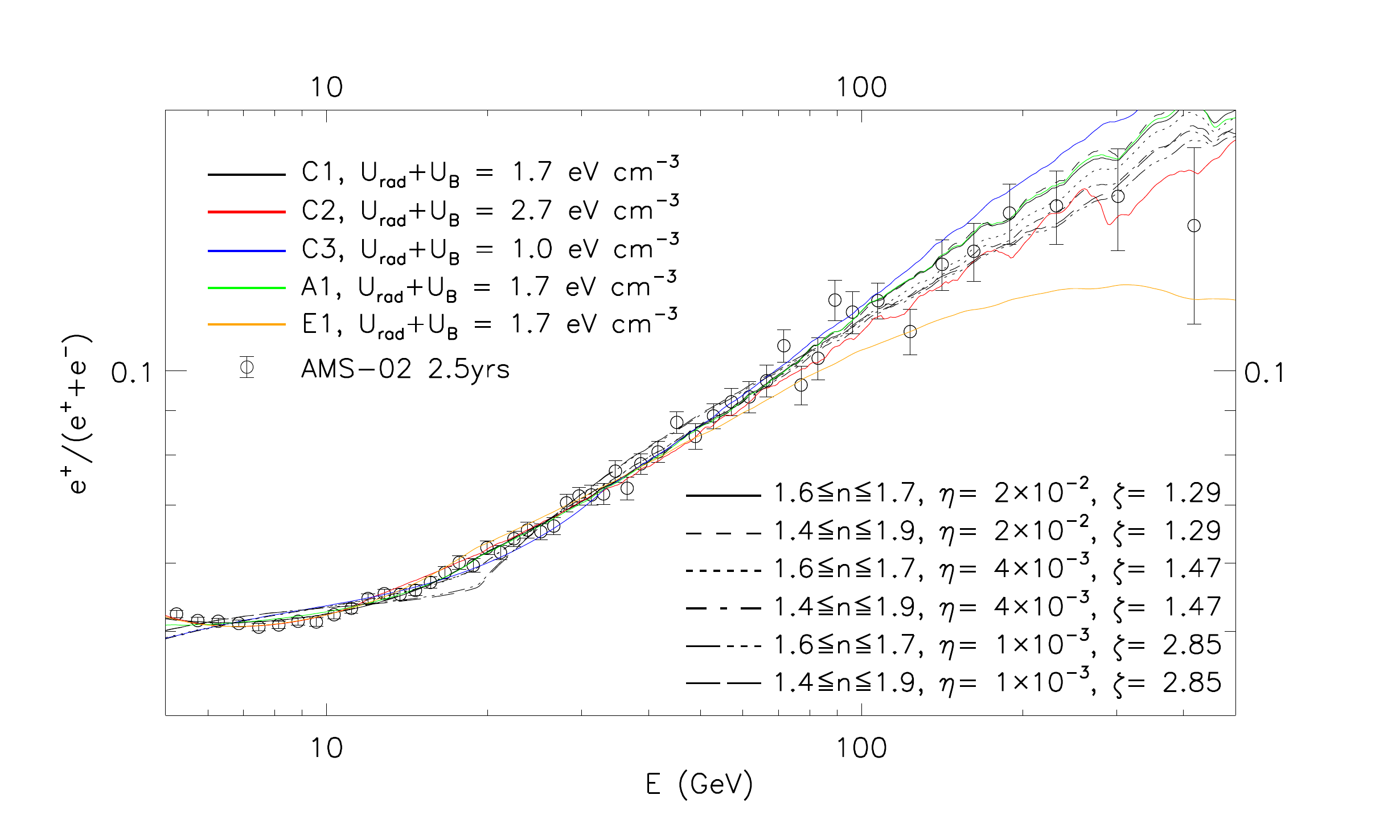}
\end{centering}
\vspace{-0.5cm}
\caption{The positron fraction, assuming pulsars follow the same distribution in space and in their spin-down power and time evolution. 
We vary the ISM propagation conditions (different colors) as in Table~\ref{tab:ISMBack}, and the CR $e^{\pm}$ spectral properties and fraction
of spin-down power into ISM injected $e^{\pm}$. $U_{\textrm{rad}}+U_{\textrm{B}}$ refer to the local energy density in the radiation and magnetic fields. 
Some of these lines are excluded in the positron fraction spectrum fit and are not used further in the PSD analysis.}
\vspace{-0.4cm}
\label{fig:ISM_Inj_Prop}
\end{figure}

The impact of these uncertainties on the morphology of the CR
spectra is shown in Figure~\ref{fig:ISM_Inj_Prop}
for the positron fraction (colored lines). Depending on the
assumptions on the energy losses and diffusion time-scales, the
spectral features can be more pronounced or suppressed. 

Once exiting the ISM and entering the heliosphere, CRs will
reach our detectors, after diffusing through the anisotropic
magnetic-field structure of the fast evolving heliospheric
magnetic field. During their propagation through the heliosphere,
CRs also transfer via 
drift effects that impact how fast they will reach Earth, and the path they 
are most prone to follow through
the Heliosphere. During that time CRs will also go through adiabatic energy losses. 
The effect of solar modulation on CR spectra is described by the
solar-modulation potential $\Phi$
that describes the average energy losses CR suffer 
as they travel through the Heliosphere. That, in terms of CR spectra, is 
given by \cite{1968ApJ...154.1011G},
\begin{eqnarray}
\frac{dN^{\oplus}}{dE_{kin}} (E_{kin}) &=& \frac{(E_{kin}+m)^{2} -m^{2}}{(E_{kin}+m+\mid Z\mid e \Phi)^{2} -m^2} \nonumber \\ 
&\times&\; \frac{dN^{\rm ISM}}{dE_{kin}} (E_{kin}+\mid Z\mid e \Phi).
\label{eq:SolMod}
\end{eqnarray}
Here, $E_{kin}$ is the kinetic energy at Earth, and 
$\frac{dN^{\oplus (\textrm{ISM})}}{dE_{kin}}$ are the
differential CR fluxes observed at Earth ($\oplus$) and the
local interstellar medium (ISM) respectively. Finally, $\mid Z
\mid e$  is the absolute charge of CRs.

Ref.~\cite{Cholis:2015gna}, using proton fluxes from 1992 and up to 2010, resulted in the predictive, time-, charge- and rigidity(R)-dependent formula for the solar modulation potential,
 \begin{eqnarray}
\Phi(R,q,t) = &\phi_{0}& \, \bigg( \frac{|B_{\rm tot}(t)|}{4\, {\rm nT}}\bigg) + \phi_{1} \, H(-qA(t))\, \bigg( \frac{|B_{\rm tot}(t)|}{4\,  {\rm nT}}\bigg) \nonumber \\
&\times& \,\bigg(\frac{1+(R/R_0)^2}{\beta (R/R_{0})^3}\bigg) \, \bigg( \frac{\alpha(t)}{\pi/2} \bigg)^{4},
\label{eq:ModPot}
\end{eqnarray}
with $R_{0}$ set to 0.5 GV and with a $2\sigma$ range for $
\phi_{0}$ of 0.32--0.38 GV and $ \phi_{1}$ in the 
range of 0--16 GV. 
We marginalize over these ranges of $ \phi_{0}$ and $ \phi_{1}$.
In Eq.~\ref{eq:ModPot} we use the values of $B_{\rm tot}(t)$ and
$\alpha(t)$ measured by ACE \cite{ACESite} and modeled in WSO
\cite{WSOSite}. Having these values, we can directly calculate
the $\Phi(R,q,t)$, for any CR species at a given rigidity and
time t. For further details see \cite{Cholis:2015gna}. 

\section{Pulsar Models Realizations versus Noise for \textit{AMS-02}}
\label{appE}

In this appendix we provide additional information on how that observation 
realizations of the pulsar models that we test perform versus the expected 
noise of \textit{AMS-02} after 20 years of measurements. 

\begin{figure}
\begin{centering}
\hspace{-0.2cm}
\includegraphics[width=3.40in,angle=0]{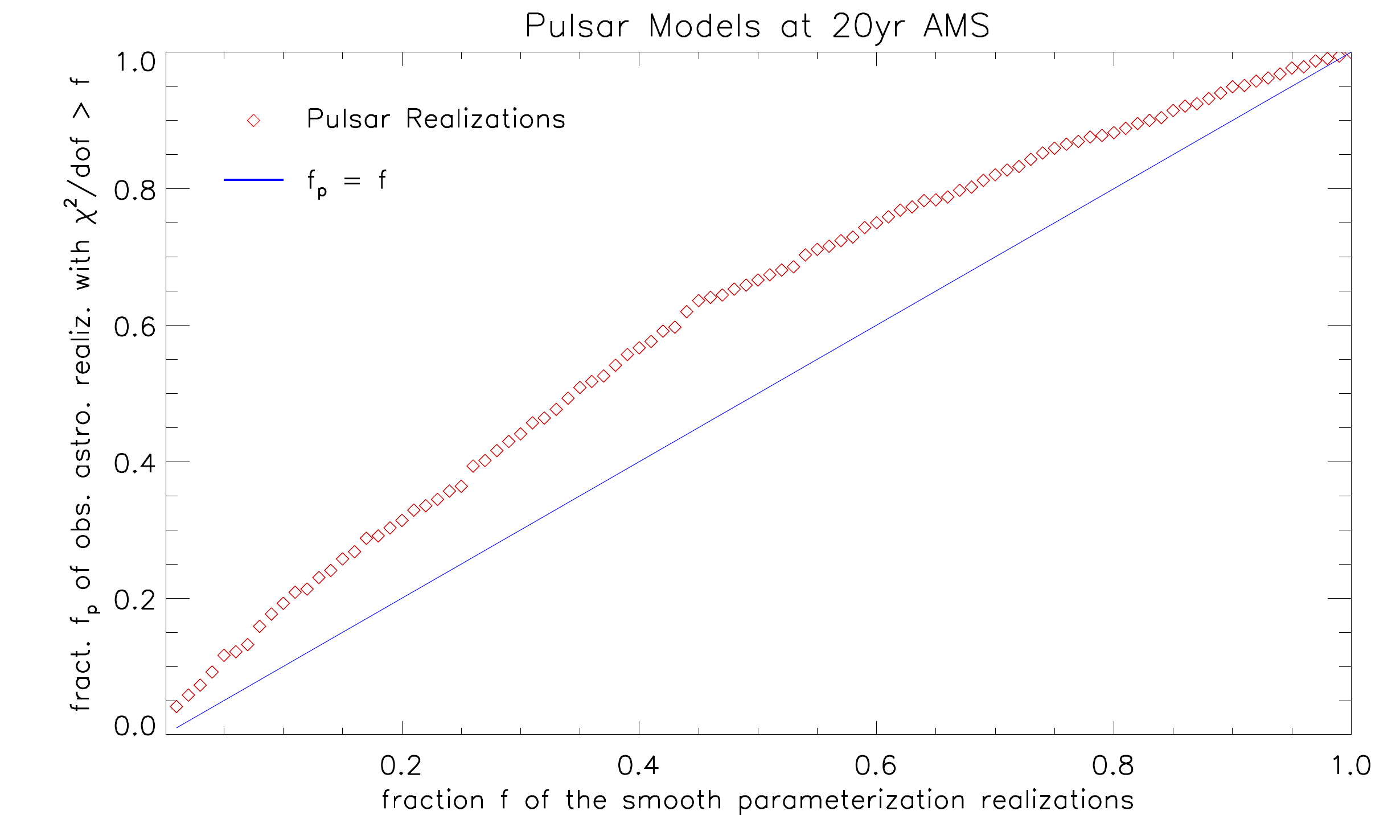}
\end{centering}
\vspace{-0.5cm}
\caption{The fraction of pulsar models observation realizations with $\chi^{2}$/d.o.f 
that is higher than the fraction $f$ of the smooth parameterization realizations, as a
function of $f$ (x-axis). The blue diagonal line gives the case where $f_{p} = f$.}
\vspace{-0.4cm}
\label{fig:Pulsar_frac_vs_Noise_frac}
\end{figure}

In Figure~\ref{fig:Pulsar_frac_vs_Noise_frac} we plot the fraction of 
observation realizations of the pulsar models  $f_{p}$ with a 
$\chi^{2}$/d.o.f  larger than $f$, vs $f$ where $f$ is the fraction 
of smooth parameterization realizations.

\end{appendix}

\end{document}